\documentclass[twoside,twocolumn,9pt]{article}
\usepackage{extsizes}
\usepackage[super,sort&compress,comma]{natbib} 
\usepackage[version=3]{mhchem}
\usepackage[left=1.5cm, right=1.5cm, top=1.785cm, bottom=2.0cm]{geometry}
\usepackage{balance}
\usepackage{mathptmx}
\usepackage{sectsty}
\usepackage{graphicx} 
\usepackage{lastpage}
\usepackage[format=plain,justification=justified,singlelinecheck=false,font={stretch=1.125,small,sf},labelfont=bf,labelsep=space]{caption}
\usepackage{float}
\usepackage{fancyhdr}
\usepackage{fnpos}
\usepackage[english]{babel}
\addto{\captionsenglish}{%
  
}
\usepackage{array}
\usepackage{droidsans}
\usepackage{charter}
\usepackage[T1]{fontenc}
\usepackage[usenames,dvipsnames]{xcolor}
\usepackage{setspace}
\usepackage[compact]{titlesec}
\usepackage{hyperref}

\usepackage{epstopdf}

\definecolor{cream}{RGB}{222,217,201}

\begin{document}

\pagestyle{fancy}
\thispagestyle{plain}
\fancypagestyle{plain}{
\renewcommand{\headrulewidth}{0pt}
}

\makeFNbottom
\makeatletter
\renewcommand\LARGE{\@setfontsize\LARGE{15pt}{17}}
\renewcommand\Large{\@setfontsize\Large{12pt}{14}}
\renewcommand\large{\@setfontsize\large{10pt}{12}}
\renewcommand\footnotesize{\@setfontsize\footnotesize{7pt}{10}}
\makeatother

\renewcommand{\thefootnote}{\fnsymbol{footnote}}
\renewcommand\footnoterule{\vspace*{1pt}%
\color{cream}\hrule width 3.5in height 0.4pt \color{black}\vspace*{5pt}} 
\setcounter{secnumdepth}{5}

\makeatletter 
\renewcommand\@biblabel[1]{#1}            
\renewcommand\@makefntext[1]%
{\noindent\makebox[0pt][r]{\@thefnmark\,}#1}
\makeatother 
\renewcommand{\figurename}{\small{Fig.}~}
\sectionfont{\sffamily\Large}
\subsectionfont{\normalsize}
\subsubsectionfont{\bf}
\setstretch{1.125} 
\setlength{\skip\footins}{0.8cm}
\setlength{\footnotesep}{0.25cm}
\setlength{\jot}{10pt}
\titlespacing*{\section}{0pt}{4pt}{4pt}
\titlespacing*{\subsection}{0pt}{15pt}{1pt}

\fancyfoot{}
\fancyfoot[LO,RE]{\vspace{-7.1pt}\includegraphics[height=9pt]{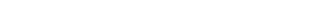}}
\fancyfoot[CO]{\vspace{-7.1pt}\hspace{13.2cm}\includegraphics{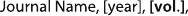}}
\fancyfoot[CE]{\vspace{-7.2pt}\hspace{-14.2cm}\includegraphics{head_foot/RF}}
\fancyfoot[RO]{\footnotesize{\sffamily{1--\pageref{LastPage} ~\textbar  \hspace{2pt}\thepage}}}
\fancyfoot[LE]{\footnotesize{\sffamily{\thepage~\textbar\hspace{3.45cm} 1--\pageref{LastPage}}}}
\fancyhead{}
\renewcommand{\headrulewidth}{0pt} 
\renewcommand{\footrulewidth}{0pt}
\setlength{\arrayrulewidth}{1pt}
\setlength{\columnsep}{6.5mm}
\setlength\bibsep{1pt}

\makeatletter 
\newlength{\figrulesep} 
\setlength{\figrulesep}{0.5\textfloatsep} 

\newcommand{\topfigrule}{\vspace*{-1pt}%
\noindent{\color{cream}\rule[-\figrulesep]{\columnwidth}{1.5pt}} }

\newcommand{\botfigrule}{\vspace*{-2pt}%
\noindent{\color{cream}\rule[\figrulesep]{\columnwidth}{1.5pt}} }

\newcommand{\dblfigrule}{\vspace*{-1pt}%
\noindent{\color{cream}\rule[-\figrulesep]{\textwidth}{1.5pt}} }

\makeatother

\twocolumn[
  \begin{@twocolumnfalse}
{\includegraphics[height=30pt]{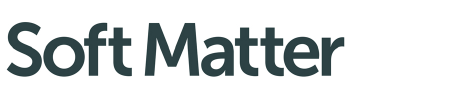}\hfill\raisebox{0pt}[0pt][0pt]{\includegraphics[height=55pt]{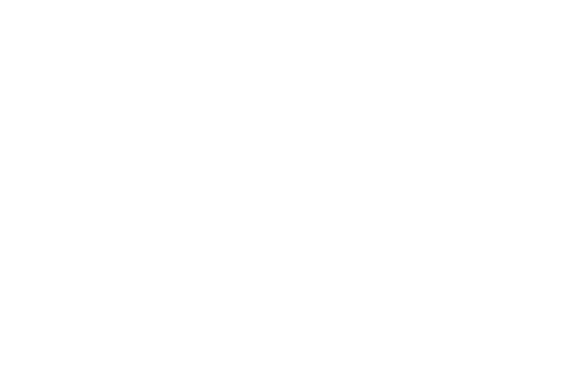}}\\[1ex]
\includegraphics[width=18.5cm]{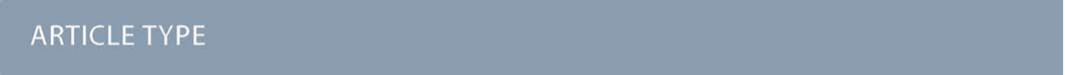}}\par
\vspace{1em}
\sffamily
\begin{tabular}{m{4.5cm} p{13.5cm} }

\includegraphics{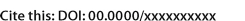} & \noindent\LARGE{\textbf{Microscopic Theory of the Elastic Shear Modulus and Length-Scale-Dependent Dynamic Re-Entrancy  Phenomena in Very Dense Sticky Particle Fluids
$^\dag$}} \\
\vspace{0.3cm} & \vspace{0.3cm} \\

 & \noindent\large{Anoop Mutneja,\textit{$^{a,d}$} and Kenneth S.Schweizer$^{\ast}$\textit{$^{a,b,c,d}$}} \\

\includegraphics{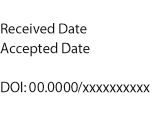} & \noindent\normalsize{{\color{black}We apply the hybrid Projectionless Dynamic Theory (hybrid PDT) formulation of the Elastically Collective Nonlinear Langevin Equation (ECNLE) activated dynamics approach to study  dense fluids of sticky spheres interacting with short range attractions. Of special interest is the problem of non-monotonic evolution with short range attraction strength of the elastic modulus (“re-entrancy”) at very high packing fractions far beyond the ideal mode coupling theory (MCT) nonergodicity boundary.} The dynamic force constraints explicitly treat the bare attractive forces that drive transient physical bond formation, while a projection approximation is employed for the singular hard-sphere potential. The resultant interference between repulsive and attractive forces contribution to the dynamic vertex results in the prediction of localization length and elastic modulus re-entrancy, qualitatively consistent with experiments. The non-monotonic evolution of the structural (alpha) relaxation time predicted by ECNLE theory with the hybrid PDT approach is explored in depth as a function of packing fraction, attraction strength, and attraction range. Isochronal dynamic arrest boundaries based on activated relaxation display the classic non-monotonic glass melting form. Comparisons of these results with the corresponding predictions of ideal MCT, and also the ECNLE and NLE activated theories based on projection, reveal large qualitative differences. The consequences of stochastic trajectory fluctuations on intra-cage single particle dynamics with variable strength of attractions are also studied. {\color{black} Large dynamical heterogeneity effects in attractive glasses are properly captured. These include  a rapidly increasing amplitude of the non-Gaussian parameter with packing fraction and a non-monotonic evolution with attraction strength, in qualitative accord with recent simulations.} Extension of the microscopic theoretical approach to treat double yielding in attractive glass nonlinear rheology is possible. } \\

\end{tabular}

 \end{@twocolumnfalse} \vspace{0.6cm}

  ]

\renewcommand*\rmdefault{bch}\normalfont\upshape
\rmfamily
\section*{}
\vspace{-1cm}


\footnotetext{\textit{~Departments of Materials Science $^{a}$, Chemistry $^{b}$, and Chemical \& Biomolecular Engineering $^{c}$, and Materials Research Laboratory $^{d}$, University of Illinois, Urbana, Illinois, 61801, USA}

$^{\ast}$kschweiz@illinois.edu


}


\section{INTRODUCTION}
The microscopic understanding of the slow activated dynamics and kinetic arrest of glass-forming fluids remains a major challenge in the fields of statistical mechanics, condensed matter physics, and materials science \cite{Angell2000,Berthier2011,Gotze2008,dhont1996}. A core idea is that particles experience prolonged local confinement in cages due to repulsive interactions and strong packing correlations, the escape from which requires activated hopping events \cite{Chaudhuri2007,Weeks2000,Miyagawa1988,Saltzman2006,Saltzman2006_2} which underlies the dramatic increase of relaxation times and viscosity \cite{Hunter2012,Berthier2011,dhont1996}. The manner in which local hopping events are correlated in space and time remains a topic of intense interest and debate.\\

{\color{black}In nanoparticle, colloidal, polymeric, and biological (e.g., proteins) soft matter, tunable strong short-range attractions can exist. Thus, long-lived physical bonds emerge, the novel dynamic consequences of which are coupled with repulsion-induced caging.} Specifically, increasing the strength of interparticle attraction can elicit, in a spatial range and fluid packing fraction dependent manner, distinctive phenomena such as re-entrant glass melting and structural relaxation \cite{Bergenholtz1999, Dawson2000, Zaccarelli2002, Pham2002, WGotze_2003, Kaufman2006, Zaccarelli2009, Willenbacher2011,Royall2018,Fullerton2020,Luo2021}, non-monotonic variation of the elastic shear modulus \cite{Pham2008, Atmuri2012,Willenbacher2011}, and a 2-step double yielding nonlinear mechanical response \cite{Pham2008, Koumakis2011, Moghimi2020, Laurati2009}. The classic qualitative picture \cite{Poon1993,Dawson2000,Pham2002,Eckert2002,Zaccarelli2002,Puertas2002,Poon2004,Royall2018} is that as the attraction strength ($\epsilon$) intensifies, bond formation first competes with repulsive cages, leading to frustration in the coherence of local cage packing and more free volume. This accelerates the dynamics, causing kinetic vitrification to occur at higher packing fractions, resulting in “re-entrancy”. {\color{black} Further increase in attraction strength leads to the formation of an attractive glass state, characterized by very strong caging and bonding which effectively reinforce. This results in an even more pronounced slowing down of dynamics compared to pure hard spheres. The corresponding kinetic arrest boundary shifts to lower packing fractions ($\phi$), imparting a distinctive nose feature (see Fig.\ref{fig:NMCT-PhaseSpace}) to the attraction strength-packing fraction ($\epsilon-\phi$) kinetic arrest map.} All these behaviors depend on the range of the attractive potential which enters, in principle, via both its effect on equilibrium structure and the magnitude and spatial variation of attractive \textit{forces.}\\

The microscopic ideal mode coupling theory (MCT) {\color{black}\cite{Gotze2008,Reichman2005,Zaccarelli2008,Zaccarelli2002,Cates2002,Pham2002,Luo2021} does predict glass melting and re-entrancy of the structural relaxation time. It can qualitatively describe the form of the kinetic arrest boundaries seen in experiment and simulation within the caveat of empirically shifting the idealized critical packing fraction for hard spheres to a significantly larger value than predicted \cite{Bergenholtz1999, Dawson2000, Zaccarelli2002, Pham2002, WGotze_2003}.} Ideal MCT predicts the relaxation time grows as an inverse critical power law, which for hard spheres diverges at $\phi=0.515$ based on Percus-Yevick (PY) integral equation input for the structure factor. The latter value is far below \cite{Hunter2012,Sciortino2005} the observed vitrification packing fraction in colloidal suspensions or in silico. The predicted power law can only describe the first few decades of the observed growth of relaxation times, which typically crosses over to a supra-exponential growth with packing fraction or attraction strength \cite{Berthier2009}. Moreover, ideal MCT predicts very weak dynamic heterogeneity effects such as decoupling of the diffusion constant and relaxation, and tiny non-Gaussian parameters on intermediate time and length scales. These limitations are generally viewed to arise from the lack of ergodicity restoring activated processes in ideal MCT, which motivated the development two decades ago of the Nonlinear Langevin Equation (NLE) theory approach \cite{Schweizer2005}. In addition, the experimental dynamic elastic shear modulus, a property determined at relatively short time and length scales that reflects the strength of localization, is also a non-monotonic function of attraction strength at high packing fractions in the attractive glass regime \cite{Pham2008,Atmuri2012}. This elastic re-entrancy is not captured \cite{Dawson2000,Luo2021} by ideal MCT, which replaces the bare repulsive and attractive forces by a single effective potential, the direct correlation function, via a projection procedure.\\

Relatively recently, a “generalized” version of ideal MCT (GMCT) has been proposed that truncates the dynamic hierarchy of time correlations functions at a higher than pair level \cite{Mayer2006,Janssen2015,Luo2021}. {\color{black} This extension is necessarily an improvement over standard ideal MCT. However, it comes at the cost of much additional complexity associated with 3 and more body correlation functions, the knowledge of which is poor.} It has been shown that truncation of GMCT at any finite level still results in a divergent relaxation time \cite{Mayer2006}, albeit it moves to larger packing fractions as more higher order many body correlations are added. Recent interesting work with this approach for the dense sticky sphere fluid \cite{Luo2021} of present interest does reduce the extent to which one must shift the ideal kinetic arrest map to higher packing fractions in order to align it with the region where the distinctive phenomena are observed in experiments and simulations. Much of the rich phenomenology associated with dynamic singularities (e.g., the $A_3$ boundary) is retained, albeit modified in detail \cite{Janssen2015,Luo2021}. However, these computationally and conceptually complex GMCT extensions do not provide a tractable route to capture activated dynamical processes in any predictive manner which requires an infinite order summation of all orders of many body correlations \cite{Mayer2006}, and which are expected to destroy the ideal MCT and GMCT singularities and special boundaries. Moreover, the ability of GMCT to capture strongly non-Gaussian or dynamically heterogeneous phenomena such as decoupling of diffusion and relaxation, large non-Gaussian parameters, and exponential tails in the van Hove function has not been demonstrated to the best of our knowledge. In contrast, the full solution of the stochastic trajectory level NLE activated dynamic theory \cite{Saltzman2006,Saltzman2006_2} does capture well the latter single particle dynamically heterogeneous features for dense hard sphere fluids \cite{Saltzman2006,Saltzman2006_2}.\\

NLE theory \cite{Schweizer2005} is a microscopic, force level, single particle statistical dynamical approach which captures thermal noise driven activated motion at the stochastic trajectory level. In NLE theory, the ideal MCT transition becomes a continuous dynamic crossover. Cage constraints, in their simplest formulation, are quantified from the pair structure, as done by MCT. The effective caging force then enters via a \textit{displacement-dependent} dynamic free energy ($F_{dyn}$), which renders the local mobility of the particles dependent on space and time. {\color{black} NLE theory has been more recently generalized to include nonlocal aspects for the long time structural relaxation. Physically, these are associated with the coupling of large amplitude particle hopping with collective elastic fluctuation of all particles outside the cage, an approach called the Elastically Collective NLE (ECNLE) theory \cite{Mirigian2014}.} ECNLE theory captures well the alpha time over the $\sim5-6$ decades of slowing down in hard sphere fluids and colloidal suspensions\cite{Mirigian2013,Mirigian2014}.\\

{\color{black} For systems with strong short range attractions, the simplified single particle (“naïve”) MCT (NMCT), NLE and ECNLE theories have recently been generalized \cite{Dell2015,Ghosh2019,Ghosh2020} to explicitly treat attractive forces. This stands in contrast to the standard projection-based approach which encodes them indirectly only via changes of the pair structure.} The resultant activated NLE and ECNLE theories predict the glass-melting effect \cite{Dell2015,Ghosh2019,Ghosh2020} under the very high packing fraction conditions they are observed in an experiment.{\color{black}  The mechanism is both the change of structure with attraction strength, and most critically the interference of dynamic cross correlations between the effective repulsive and attractive forces when the latter is of intermediate magnitude.} This basic mechanism of re-entrancy is qualitatively different than in ideal MCT where glass melting is typically attributed to the non-monotonic dependence with attraction strength of the collective structure factor amplitude near its cage peak, $S(q^\ast)$ \cite{Ghosh2019}. Evidently, the latter non-monotonic behavior does not extend to large packing fractions ($\phi$) where elastic modulus and structural relaxation \cite{Fullerton2020,Luo2021} behavior is experimentally observed. 
{\color{black} Moreover, as mentioned above, very strong single particle signatures of dynamic heterogeneity in attractive glasses on intermediate (“in cage”) time and length scales are not captured by MCT  \cite{Zaccarelli2008,Hunter2012,Weeks2000,Chaudhuri2007,Luo2021}. To date, they also have not been analyzed based on NLE or ECNLE theory formulated at the level of explicit attractive forces.}\\

{\color{black}The need for a new theoretical approach to the above problems is buttressed by the recent simulation study of Fullerton and Berthier \cite{Fullerton2020}.These workers employed swap Monte Carlo to numerically extend the study of the typical re-entrancy features to very large packing fractions deep in the attractive glass regime.} A nose-like isochronal kinetic arrest boundary is observed, but far from any ideal MCT non-ergodicity transition, and far from where $S\left(q^\ast\right)$ is non-monotonic. A non-monotonic behavior of the Debye-Waller factor with attraction strength was also found, suggestive of the experimentally observed elastic modulus re-entrancy \cite{Pham2008,Atmuri2012}. Moreover, systems above and well beyond the nose (attractive glasses) exhibited stronger dynamic heterogeneity on intermediate length scales. Overall, these workers argued the singularities predicted by MCT, which, for example, leads to the so-called logarithmic decay of the dynamic structure factor at high packing fractions, cannot be employed to understand the rich dynamics of ultra-dense attractive particle fluids in the regime where activated motion is clearly of paramount importance. \\

Although not the focus of the present work, we mention in passing that the nontrivial, but rather subtle, dynamical differences between a Lenard-Jones (LJ) fluid and its purely repulsive counterpart, the Weeks-Chandler-Andersen (WCA) fluid, is apparently not captured by ideal MCT \cite{Berthier2010} since these two systems have very similar structure factors. The origin of this failure remains debated, including the possibility higher structural correlations are important and other diverse ideas \cite{Tong2020,Landes2020,Nandi2017,Berthier2011_2,Coslovich2007,Chen2010,Jacques2017}. {\color{black} This difference has been argued to be understandable based on ECNLE theory extended to treat attractive forces explicitly and avoid a literal projection approximation \cite{Dell2015}. This approach is referred to as the Projectionless Dynamic Theory (PDT) \cite{Schweizer1982, Schweizer1989} formulation of NMCT, NLE, and ECNLE theories. }\\

In this study, we adopt a hybrid version of the PDT in the ECNLE theory framework \cite{Dell2015,Ghosh2019,Ghosh2020} to address the following topics in ultra-dense sticky particle fluids: (i) a much broader and deeper exploration of the packing fraction, attraction strength, and attraction range dependence of structural relaxation and kinetic arrest diagrams, (ii) construction of the first understanding of elastic shear modulus re-entrancy in attractive glasses and comparision to experiment, and  (iii) application of the full stochastic trajectory version of NLE theory to study for the first time the time-dependent mean square displacement (MSD) and non-Gaussian effects in the attractive glass regime associated with stochastic trajectory fluctuations on intra-cage length scales.\\

In section {\ref{Section2}}, we briefly review the underlying theoretical approaches employed. New results are presented in section \ref{Section3} which contrast the projected and hybrid PDT predictions of NMCT, and new predictions are made for the elastic shear modulus and dynamic localization length re-entrancy. Section \ref{Section4} analyzes all key length scales of the dynamic free energy which play a crucial role for thermally activated dynamics, and presents results based on the Brownian simulation solution of the NLE evolution equation for the MSD and NGP of attractive dense fluids. Section \ref{Section5} studies the non-monotonic structural relaxation times and constructs isochronal kinetic arrest maps in the strongly activated regime. The article concludes in section \ref{Section6} with a discussion and outlook to future directions that the present work facilitates.

\section{THEORETICAL BACKGROUND}\label{Section2}
\subsection{Projected versus Projectionless Formulation of Dynamic Constraints }
In single particle (naïve) Mode-Coupling Theory (NMCT) and its beyond MCT extensions, the force-force time correlation function, $K\left(t\right)=\frac{\beta}{3}\left\langle\vec{F_0}\left(0\right).\vec{F_0}\left(t\right)\right\rangle$ is the starting point. Here, $\vec{F_0}\left(t\right)$ is the total force at time $t$ on a tagged particle due to its surroundings. In classic MCT-like approaches, the slow force dynamics is assumed to be governed by structural pair correlations, with real forces projected onto slow bilinear density modes regardless of how many different types of forces exist at a microscopic level. While this is a benign simplification for the long wavelength classic problem of critical slowing down,  or purely repulsive force systems such as hard spheres, its accuracy for local dynamics involving competing attractive and repulsive forces is unclear. After projection, one has a four-point dynamic correlation function which is then factorized into the product of two-point functions in a dynamically Gaussian manner. In Fourier space, this factorization yields:
\begin{equation}\label{eqn:memory_function}
        K(t)=\frac{1}{3}\beta^{-1}\int \frac{d\textbf{q}}{(2\pi)^3} |\vec{M}(q)|^2 \rho S(q)\Gamma_S(q,t)\Gamma_C(q,t).    
\end{equation}
Here, $\rho$ is the particle number density, $S\left(q\right)$ is the dimensionless collective structure factor, and $\Gamma_S\left(q,t\right)=\left\langle e^{i\vec{q}\cdot\left(\vec{r}\left(t\right)-\vec{r}\left(0\right)\right)}\right\rangle$ and $\Gamma_C\left(q,t\right)=S\left(q,t\right)/S\left(q\right)$ are the normalized single particle and collective dynamic structure factors, respectively. The so-called force vertex $\vec{M}\left(q\right)$ in Eq.~(\ref{eqn:memory_function}) replaces the real forces by pair structure information :
\begin{equation}\label{eqn:Projection}
    \vec{M}_{NMCT}(q)=qC(q)\hat{q}.
\end{equation}
where $C\left(q\right)=\rho^{-1}\left[1-S^{-1}\left(q\right)\right]$ is the direct correlation function, and $S\left(q\right)$, $C\left(q\right)$, and the Fourier transform of the non-random part of the pair correlation function, $h\left(q\right)$, are related via $S\left(q\right)=1+\rho h\left(q\right)$ and $h\left(q\right)=C\left(q\right)S\left(q\right)$.This dynamical scheme will be referred to as the “Projected” theory.\\

A projectionless dynamical theory (PDT) \cite{Schweizer1982,Schweizer1989} was more recently formulated\cite{Dell2015,Ghosh2019} which explicitly retains the true pairwise decomposable forces to construct dynamic constraints. The effective force is given by:
\begin{equation}
    \begin{split}
        \vec{M}_{PDT}(q)&={\color{black}\beta}\int d\vec{r}\vec{f}(r)g(r)e^{-i\vec{q}.\vec{r}}\\
        &=4\pi{\color{black}\beta}\hat{r}\int_0^\infty r^2f(r)g(r)\frac{\sin(qr)}{qr}dr,
    \end{split}
\end{equation}
where $f\left(r\right)$ is the bare interparticle force. The PDT approach as embedded in NLE and ECNLE activated dynamics theories for pure hard-spheres has been shown to qualitatively capture the slowing down in colloidal experiments and hard sphere simulations \cite{Dell2015}, but tends to quantitatively overpredict the activation barrier. To address this issue, a hybrid approach was proposed \cite{Dell2015,Ghosh2019}, where the singular repulsive hard-sphere potential was handled using the usual projected method,  while the attractive force was treated in a projectionless manner. This strategy follows a long history in chemical physics \cite{Schweizer1982,Schweizer1989,Hansen2006-pc} of acknowledging the strong differences between the dynamical consequences of repulsive and attractive forces, and their treatment with different physical ideas. For a system with a hard core repulsion and short-range attraction described by Eq.~(\ref{eqn:Potential}), the hybrid effective force is \cite{Dell2015,Ghosh2020_2}:
\begin{equation}\label{eqn:HybridForceVertex}
    \begin{split}
        \vec{M}_{Hyb}(q)&\approx\vec{M}_R(q)+\vec{M}_A(q)\\
        &=\vec{M}_{NMCT}(q)+\vec{M}_{PDT}(q)\\
        &=qC_0(q)\hat{q}+4\pi{\color{black}\beta}\hat{r}\int_0^\infty r^2f(r)g(r)\frac{sin(qr)}{qr}dr,\\
    \end{split}
\end{equation}
The repulsive contribution (first term) arises from the hard sphere system, while the attractive force is directly treated. In the present study, we adopt the following pair potential
\begin{equation}\label{eqn:Potential}
   V(r)= 
\begin{cases}
    \infty,& r\leq \sigma\\
    -\epsilon e^{-\frac{r-\sigma}{a}},& r>\sigma
\end{cases}
\end{equation}
where $\epsilon$ denotes the attraction strength at contact, and $a$ is its spatial range. {\color{black}As widely discussed in the theoretical and simulation community, we do not expect any of our core findings are substantively affected by the precise form of the attractive potential as long as it is short range with a well defined length scale.} Unless stated otherwise, dimensional energy is presented in units of $\beta^{-1}=k_BT$, and dimensional length is in units of the particle hard core diameter, $\sigma$.\\

The above is the main working theory adopted in this article, and will be referred to as the hybrid PDT. Note that we are not using any wavevector-dependent switching between the projection and projectionless theories as was discussed in ref.~\cite{Ghosh2019}. The dynamical force vertex includes a negative interference or cross term between the attractive (negative) and repulsive (positive) contributions, which can reduce the net dynamical constraints. Consequently, for small attractions, the hard sphere constraints are reduced, leading to glass melting \cite{Ghosh2019}. As attractions become stronger, the diagonal square term associated with strong physical bonding starts to dominate, enhancing the dynamic constraints. This leads to a non-monotonic variation of dynamic and elastic properties at high packing fractions. This explicit force effect is not present in theories based on projection, where attractive interactions only enter as they modify the pair structure. The next section concisely reviews the basic aspects of the NMCT, NLE, ECNLE approaches.

\subsection{Naïve Mode Coupling Theories~~} NMCT predicts ideal kinetic vitrification via a self-consistent dynamic closure at the most elementary mean squared displacement (MSD) level. A Vineyard-type \cite{Hansen2006-pc} approximation including the \cite{Schweizer2005,Ghosh2023} deGennes narrowing effect is employed to connect the collective dynamic structure factor and its single-particle counterpart.
\begin{equation}\label{eqn:Vineyard}
    \Gamma_c(\textbf{q},t)\approx\Gamma_s\left(\textbf{q}/\sqrt{(S(\textbf{q}))},t\right)
\end{equation}
The very large importance of the deGennes correction to the collective dynamic structure factor in NMCT \cite{Saltzman2003}, and especially for activated dynamics (NLE, ECNLE levels), has been recently established by Ghosh \cite{Ghosh2023}. A Gaussian approximation for the arrested solid state amorphous structure is adopted \cite{Saltzman2003,Kirkpatrick1987} corresponding to a harmonic Einstein glass characterized by a single dynamical localization length, $r_L$:
\begin{equation}\label{eqn:GaussianApprox}
    \Gamma_s\left(\textbf{q},t\rightarrow{}\infty\right)=e^{-\frac{q^2r_L^2}{6}}
\end{equation}
The localization length follows from the derived self-consistent equation \cite{Schweizer2005} 
\begin{equation}\label{eqn:localization_length}
    \frac{1}{r_L^2}=\frac{1}{9}\int\frac{\vec{dq}}{(2\pi)^3}\rho |\vec{M}(q)|^2 S(q) e^{-\frac{q^2r_L^2}{6}[1+S^{-1}(q)]}.
\end{equation}
The first emergence of finite value of the localization length indicates particle localization and the ideal kinetic arrest transition. \\

All theories require accurate input for the structure factor, typically from integral equation theory. While the PY \cite{Hansen2006-pc} closure is frequently used for hard and sticky spheres, recent studies, including comparison to simulations in the deeply metastable regime of central interest in this work, indicate that the modified Verlet (MV) \cite{Verlet1980,Zhou2020} closure provides significantly more accurate structural correlations at the high packing fractions relevant to strongly activated dynamics \cite{Zhou2020}. Thus, we adopt this closure.  One finds an ideal NMCT kinetic arrest at $\phi_c=0.44$, essentially the same as found based on the PY closure since this packing fraction is not in a deeply metastable regime.\\

Adopting the philosophy that slow stress fluctuations and density fluctuations are strongly coupled, the elastic shear modulus ($G^\prime$) at the NMCT level follows as \cite{Ngele1998}:
\begin{equation}\label{eqn:GprimeMCT}
    G^\prime=\frac{k_BT}{60 \pi^2}\int_0^\infty dq \left[q^2 \rho S(q) \frac{d C(q)}{dq}\right]^2\exp\left(-\frac{q^2r_L^2}{3S(q)}\right)
\end{equation}
Here, $C\left(q\right)$ is proportional to the Fourier transform of the effective force experienced by a tagged particle in a literal projection formulation. For the hybrid PDT approach that explicitly distinguishes repulsive and attractive forces, and hence for consistency with Eqs~(\ref{eqn:memory_function}) and (\ref{eqn:HybridForceVertex}), one has
\begin{equation}{\label{eqn:Gprime}}
    G^\prime=\frac{k_BT}{60 \pi^2}\int_0^\infty dq \left[q^2 \rho S(q) \frac{d (|\vec{M}(q)|/q)}{dq}\right]^2\exp\left(-\frac{q^2r_L^2}{3S(q)}\right)
\end{equation}
with $M\left(q\right)$ given in Eq.(\ref{eqn:HybridForceVertex}).
\subsection{Activated Dynamical Theories~~}
To go beyond ideal NMCT, the NLE theory was formulated using dynamic density functional ideas at the non-ensembled averaged level \cite{Schweizer2005} to construct a stochastic nonlinear evolution equation for the dynamical displacement of a particle from its initial position. In the overdamped limit of interest, three forces enter: a short time and distance frictional drag, the corresponding fluctuating random force, and a particle-displacement-dependent effective caging force on a tagged particle due to all the particles denoted as $-\partial F_{dyn}/\partial r$, where $F_{dyn}$ is the dynamic free energy. The force-balance evolution equation for the angularly averaged scalar displacement of a tagged particle is :
\begin{equation}\label{eqn:NLE}
-\zeta_s\frac{dr}{dt}-\frac{\partial F_{dyn}}{\partial r}+\delta f=0
\end{equation}
Here, $\zeta_s$ denotes the friction constant associated with non-activated very local and short time dissipative processes, and the thermal noise obeys $\left\langle\delta f\left(0\right)\delta f\left(t\right)\right\rangle=2k_BT\zeta_s\delta\left(t\right)$. In this article, time will be expressed in terms of the corresponding short time scale $\tau_s=\beta\zeta_s\sigma^2$, where the explicit expression for spheres can be found in ref. \cite{Schweizer2005}. Without a noise term, the NLE equation reduces to the self-consistent NMCT localization length equation where its solution corresponds to the minimum of $F_{dyn}$. In the presence of ergodicity restoring thermal fluctuations, the dynamic free energy is \cite{Schweizer2005}.
\begin{equation}\label{eqn:Fdyn}
\begin{split}
\beta F_{dyn}(r)=-3\ln(r)-\frac{\rho}{2\pi^2}\int_0^\infty\frac{|\vec{M}(q)|^2S(q)}{1+S^{-1}(q)}e^{-\frac{q^2r^2}{6}(1+\frac{1}{S(q)})}dq.
\end{split}
\end{equation}
A typical dynamic free energy is shown in Fig.~\ref{fig:FDyn}(a). Beyond the NMCT ideal arrest transition, a local barrier $F_B$ emerges at $r=r_B$. One can determine the mean barrier hopping (barrier crossing) time using the Kramers formula as a proxy for the mean alpha relaxation time:
\begin{equation}\label{eqn:TauAlpha}
\frac{\tau_\alpha^{hop}}{\tau_s}=\int_{r_L}^{r_B}dx~e^{\beta F_{dyn}(x)}\int_{r_L}^x dy~e^{-\beta F_{dyn}(y)}
\end{equation}\\

NLE theory has been extended \cite{Mirigian2014} to account for the effects of out-of-cage elastic distortions associated with large amplitude local hopping events on the activated alpha relaxation time, an approach called ECNLE theory. It is based on the idea that a cage escape event requires, or is facilitated by, a spontaneous elastic distortion of all the surrounding particles, resulting in an additional elastic barrier \cite{Dyre1996,Dyre1998,Dyre2006}. The elastic barrier is calculated within the Einstein glass framework as $F_{el}=4\pi\int_{r_{cage}}^{\infty}r^2\rho g\left(r\right)\left(\frac{1}{2}K_0u\left(r\right)^2\right)dr$, where $K_0$ is the harmonic spring constant of $F_{dyn}$ at its minima, and $u\left(r\right)$ is the displacement field required for a cage escape with $r$ the distance from the centre of the cage. The displacement field is constructed in the spirit of continuum elasticity \cite{Dyre2006}, since the required elastic displacements are predicted (not assumed) to be very small, $u\left(r\right)=\Delta r_{eff}\left(\frac{r_{cage}}{r}\right)^2$ for $r\geq r_{cage}$, and $r_{cage}$ is identified as the distance at the first minimum of $g\left(r\right)$. The amplitude ($\Delta r_{eff}$), or effective cage expansion, follows from a microscopic analysis of the mean extent to which cage scale hopping results in a particle displacement larger than the cage size. Defining the microscopic jump distance $\Delta r=r_B-r_L$, this analysis yields \cite{Mirigian2014}:
\begin{equation}\label{eqn:ECNLE_cageExpansion}
\Delta r_{eff}\approx\frac{3}{r^3_{cage}}\left(\frac{r^2_{cage}\Delta r^2}{32} -\frac{r_{cage}\Delta r^3}{192} +\frac{\Delta r^4}{3072} \right)
\end{equation}
The alpha time for the 2-barrier based ECNLE theory follows by introducing the extra multiplicative factor $e^{\beta F_{el}}$  in Eq.~(\ref{eqn:TauAlpha}). The final expression for the mean alpha time for systems where activation barriers are predicted is: \\
\begin{equation}\label{eqn:TauAlphaEl}
\tau_\alpha = \tau_\alpha^{hop}e^{\beta F_{el}}.    
\end{equation}

Importantly, the elastic barrier requires only information contained in the dynamic free energy. ECNLE theory predicts at sufficiently low packing fractions or high temperatures for thermal liquids, the elastic barrier is unimportant, and the activated event is local \cite{Mirigian2014}. However, with sufficient densification or cooling, since the elastic barrier is predicted (not assumed) to grow with these control variables faster than its local cage analog,  eventually the elastic barrier becomes of paramount importance in the determination of the alpha time. Hence, the alpha relaxation process becomes a coupled local-nonlocal character.  The precise manner this interplay of barriers evolves is system-specific \cite{Xie2016,Mei2020}.\\

To summarize, the hybrid PDT-based version of ECNLE theory employed in this work differs from conventional ideal MCT for sticky particles in the following respects: (1) explicitly treats attractive interactions via the hybrid force vertex (Eq.~(\ref{eqn:HybridForceVertex})) in the calculation of all dynamical properties, (2) thermally-induced activated motion is included for all “uphill” processes on the dynamic free energy, including intermediate length scale “in cage” displacements, (3) the longer range collective motion of particles outside the cage is included, which is generally critical for the larger displacement associated with barrier crossing and the alpha time, but not for smaller displacements that control the elastic modulus and time-dependent MSD and non-Gaussian trajectory fluctuation effects inside the cage. Point (1) will be demonstrated below to be crucial for understanding the elastic modulus re-entrancy effect. Points (1) and (2) are essential for understanding intermediate time and length scale dynamical phenomena. Points (1)-(3) are critical for slow activated relaxation, and constructing isochronal kinetic arrest maps for the emergence of a solid on the experimental timescale. We emphasize there are no divergences in the theory above zero Kelvin or below RCP, and the NMCT ``transition” becomes a smooth crossover to an activated dynamical regime ruled at the trajectory level by the dynamic free energy. All calculations employ the MV closure for structural input \cite{Zhou2019}. Finally, comparison with elementary aspects of conventional ideal NMCT framework are achieved by omitting the noise term and utilizing the projected force vertex.
\section{DYNAMIC LOCALIZATION AND ELASTIC SHEAR MODULUS}\label{Section3}
\subsection{Ideal Kinetic Arrest Boundary}
\begin{figure}
    \includegraphics[width=0.40\textwidth]{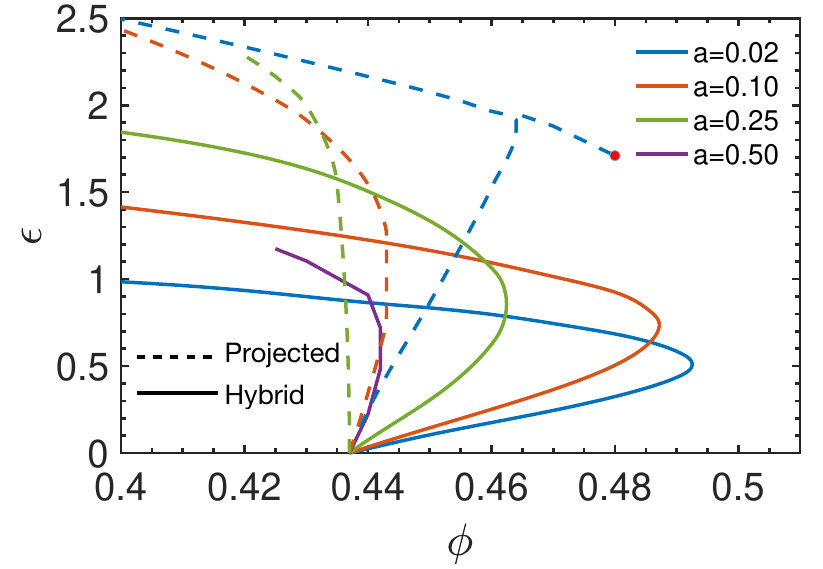}
    \caption{Ideal kinetic arrest map in attraction strength (in units of thermal energy) versus packing fraction space based on NMCT with the projected effective force (dashed curves) and its hybrid projectionless analog (solid lines) for four different short ranges of exponential attractive forces. The $A_3$ singularity line is also shown for projected, $a=0.02$ case, while it does not exist in the hybrid theory.}
    \label{fig:NMCT-PhaseSpace}
\end{figure}

Figure~\ref{fig:NMCT-PhaseSpace} shows the NMCT ideal kinetic arrest boundaries as solid curves based on the hybrid-PDT dynamic force vertex, for several choices of attraction ranges. A non-monotonic re-entrant behaviour is predicted, which decreases in amplitude with increasing attraction range. The dimensionless energy scale of the “nose” feature lies in the range of $0.5-1.0$ and varies with attraction range. The corresponding NMCT arrest boundaries using the projected vertex are shown as the dashed curves. Key differences include: the non-monotonic aspect is weaker, decreases more quickly with growing attraction range, and the attraction strength at the nose is larger. These differences reflect the explicit treatment of attractive forces and the presence of an interference term between repulsive and attractive forces in the hybrid PDT approach absent in its projected analog. Very importantly, we find the $A_3$ line \cite{Dawson2000,Priya2014} extending to the right of the nose feature predicted by the projected NMCT theory is \textit{not} present for the hybrid PDT calculations. This result seems consistent with the findings of recent simulations \cite{Fullerton2020} that challenge the existence of this feature.
\subsection{Localization Length and Shear Modulus}
\begin{figure}
    \centering
    \includegraphics[width=0.39\textwidth]{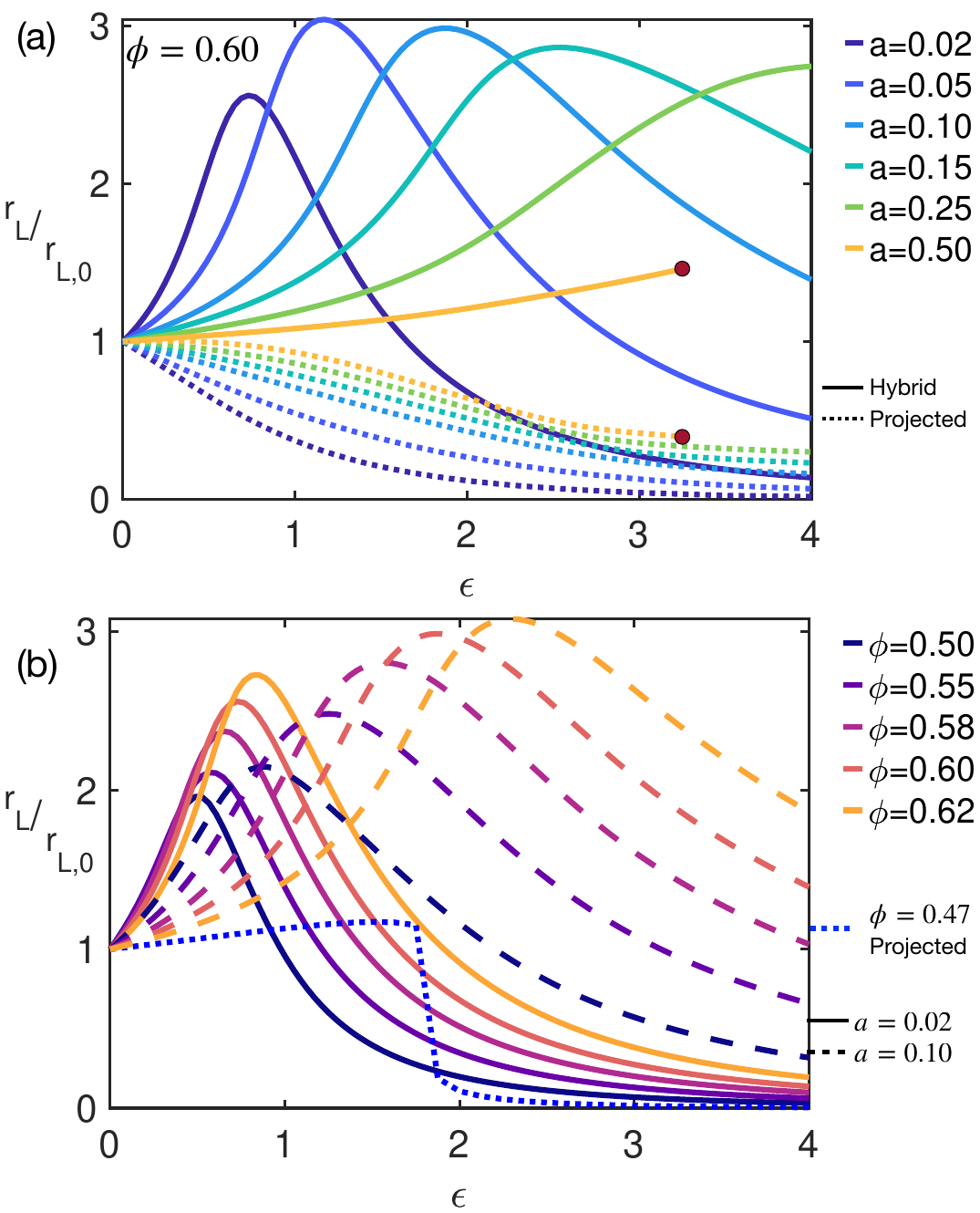}
    \caption{(a) Localization length normalized to its corresponding hard sphere fluid value as a function of attraction strength for different attraction ranges at a fixed value of packing fraction $\phi=0.60$. {\color{black}Solid (dashed) curves indicate prediction based on the hybrid-PDT (projected) vertex.} The red point marks the spinodal point. (b) Attraction amplitude dependence of localization length using the hybrid-PDT theory for different packing fractions. Solid lines are for attraction range $a=0.02$, while the dashed lines are for $a=0.10$. {\color{black} The $\phi=0.47$ dotted curve obtained using the projected theory shows a discontinuous jump per a $A_3$ singularity.} }
    \label{fig:RLoc}
\end{figure}

In NMCT the dynamic order parameter is the localization length $r_L$. Figure~\ref{fig:RLoc}(a) illustrates its variation at a packing fraction $\phi=0.60$ based on the hybrid-PDT approach (solid curves) and the projected calculation (dotted curves). Large qualitative differences are seen. Within the projected description, the localization length monotonically decreases with attraction strength, per the typical MCT behavior \cite{Dawson2000} (in terms of Debye-Waller factor) at high packing fractions. In contrast, the hybrid-PDT based theory predicts a non-monotonic behavior where the localization length initially increases with attraction strength, and then decreases, and is of order the attraction range at high attraction strengths. Such non-monotonic behaviour is supported by simulation studies at high packing fractions \cite{Fullerton2020} {\color{black} in terms of the finding of a non-monotonic Debye-Waller factor which indicates non-monotonicity of the dynamic localization length and plausibly the dynamic plateau shear modulus}, and is not explained by MCT based on the standard projection approximation \cite{Dawson2000}. This difference will be shown to have huge consequences for all dynamical properties at both the NMCT and activated NLE and ECNLE theory levels. \\

We note that the common operational approach of shifting the ideal MCT predictions based on literal singularities at packing fractions lower than relevant to experiment or simulation to higher-$\phi$ yields a discontinuous jump in the localization length associated with the $A_3$ singularity \cite{Dawson2000,Luo2021}. However, as mentioned above, recent simulations \cite{Fullerton2020} suggest the latter does not exist in a high packing fraction attractive glass regime. A related point is within the projection-based NMCT the localization length does exhibit a discontinuous jump per a $A_3$ singularity close to the NMCT phase boundary (see Fig.~\ref{fig:RLoc}(b) for $\phi=0.47$), but, crucially, not in the very dense regime of prime interest in experiments and simulations.\\

Fig.~\ref{fig:RLoc}(a)  also illustrates the non-monotonic behavior of the localization length for various attraction ranges, $a\in\left[0.02-0.50\right]$, with an amplitude that varies non-monotonically with range. The suppression of the amplitude for the longest range attraction, $a=0.5$ (calculations limited by the presence of a spinodal), seems consistent with the idea attractive forces have little dynamical consequences at high concentrations per the classic van der Waals paradigm \cite{Chandler1983}. Figure~\ref{fig:RLoc}(b) shows the localization length variation with dimensionless attraction strength, $\epsilon$, for different packing fractions scaled by the hard sphere value for two spatial ranges $a=0.02$ (solid lines) and $a=0.10$ (dashed lines). The magnitude of the non-monotonic behaviour increases at higher  $\phi$, and also requires a larger attraction strength to achieve the maximum glass melting state.\\

\begin{figure}
    \centering
    \includegraphics[width=0.45\textwidth]{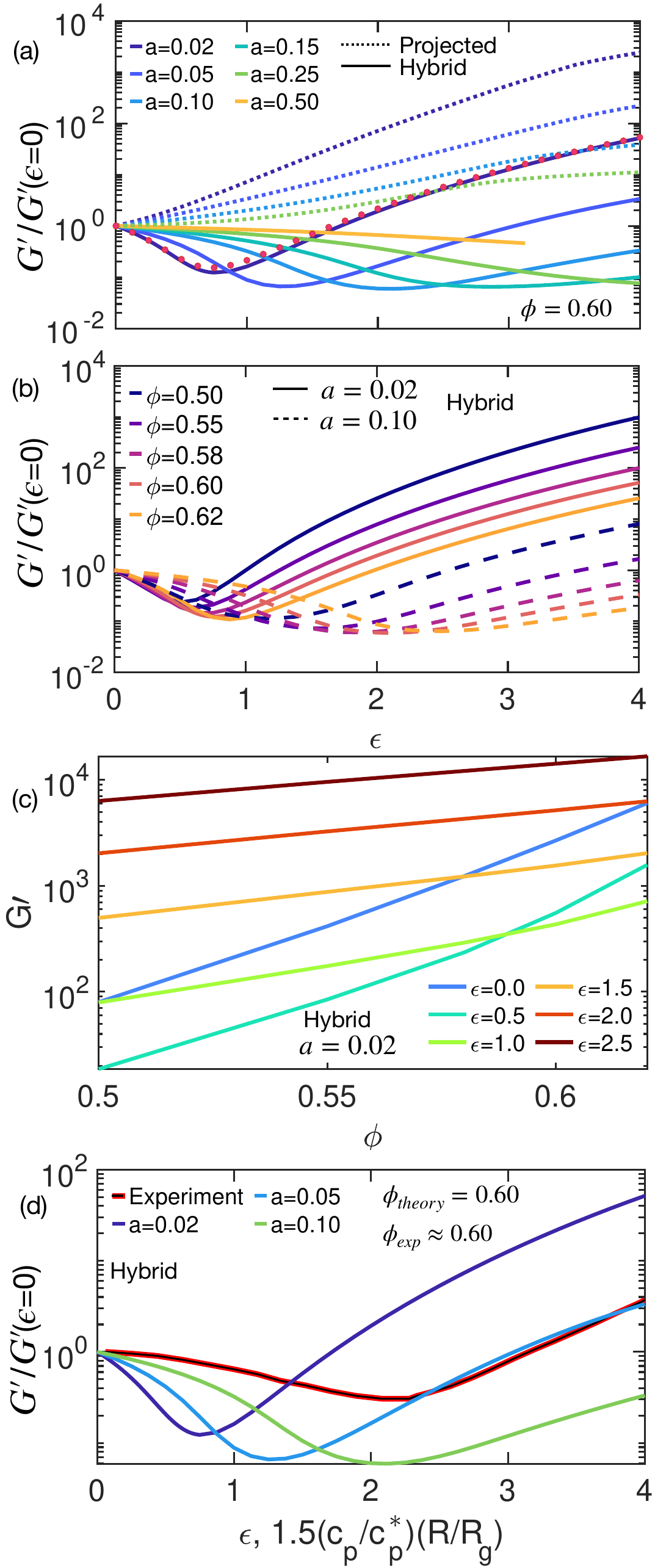}
    \caption{(a) {\color{black} Shear modulus normalized by its hard sphere value as a function of attraction strength for different attraction ranges at a fixed value of packing fraction $\phi=0.60$ using the hybrid PDT vertex (solid), and the projected vertex (dotted).} The red points compare the shear modulus with the micro-rheological relation, $G^\prime\sim \frac{k_BT\phi}{r_L^2 \sigma}$. (b) Evolution of {\color{black} scaled} shear modulus with attraction strength for different packing fractions. Solid lines are for attraction range $a=0.02$, while the dashed lines are for $a=0.10$. (c) Growth of the {\color{black}dimensionless} shear modulus with packing fraction for repulsive and attractive glasses. (d) The elastic re-entrancy predicted by the theory for the 1-component fluid with exponential attractions of three ranges is compared with the experiments of ref.~\cite{Pham2008} on colloid-polymer mixtures.}
    \label{fig:GPrime}
\end{figure}
Using Eq.~(\ref{eqn:Gprime}), the elastic shear modulus has been determined. Figure~\ref{fig:GPrime}(a) illustrates its dependence on attraction strength for various attraction ranges at a fixed high packing fraction. An elastic re-entrant behavior is predicted based on the hybrid PDT vertex. The modulus first decreases with increasing attraction strength, followed by an eventual increase due to the formation of strong bonds. The elastic shear modulus is also favorably compared with the theoretically derived micro-rheological relation $G^\prime\sim \frac{k_BT\phi}{r_L^2 \sigma}$ \cite{Kobelev2005,Mirigian2014} in terms of the inverse localization length squared (shown as red points), which provides a simple physical interpretation of the full numerical results.
The plot shows the starkly different NMCT results using the projected vertex, where the modulus increases monotonically with attraction strength. The packing fraction dependence of these behaviors is shown in Fig.~\ref{fig:GPrime}(b). The non-monotonic evolution is very similar to that of the localization length, as expected. Figure~\ref{fig:GPrime}(c) illustrates the growth of $G^\prime$ with packing fraction for different attraction strengths at a fixed short range.  The modulus of the repulsive glass increases in a rapid and elastically fragile exponential manner as previously discussed \cite{Rao2006}, while the attractive glass has a higher modulus but exhibits a weaker response to packing fraction akin to an elastically strong behavior.\\

Overall, we again emphasize that the rich variation with attraction range in Figs.~1-3 (including some non-monotonic variations), and the differences compared to NMCT predictions with a projected vertex, reflect both the change of pair structure as it enters the force vertex via $g\left(r\right)$ in the hybrid theory, and the direct explicit effect that the bare attractive force increases as $\epsilon/a$ and there are cross correlations between the different sign attractive and repulsive forces. We note that based on the recent predictions of the non-ergodocity parameter of the GMCT approach for sticky particle fluids \cite{Luo2021}, we expect that the non-monotonic elastic modulus behavior is also not captured at this higher level of MCT based on force projection.\\

Figure~\ref{fig:GPrime}(d) contrasts the theoretical results for the elastic re-entrancy effect with experiments on colloid-polymer mixtures \cite{Pham2008} at a high colloidal packing fraction of $\phi\approx0.6$. The experimental data is extracted from the frequency-dependent storage modulus obtained from small amplitude oscillatory shear measurements in the linear regime as a function of polymer concentration which tunes the entropic depletion attraction strength. The specific system studied consists of sterically stabilized poly(methylmethacrylate) (PMMA) colloids with a hydrodynamic radius ($R$) of $130 nm$, mixed with dilute solutions of varying concentrations ($c_p$) of the non-adsorbing polystyrene polymer of a radius of gyration ($R_g$) of $11 nm$. Crudely,  at the pair potential level (an approximation \cite{Chen2004,Chen2005}), the depletion attraction can be described by the simplified Asakura Oosawa potential  \cite{Asakura1954,AOPotentialJCP2022} which models polymers as small ``phantom'' spheres that do not interact with each other, but interact with the larger colloidal particles as hard spheres. The dimensionless range of this potential is estimated as the ratio $R_g/R$ ($\approx0.084$), while the attraction strength has the magnitude of \cite{Chen2004} $\sim\frac{3}{2}\frac{c_p}{c_p^\ast}\frac{R}{R_g}$. Here $c_p^\ast$ is the dilute-to-semidilute overlap concentration, and $c_p/c_p^\ast$ is varied in range $\in\left[0-0.238\right]$.\\

To compare with the theory results based on a 1-component monodisperse fluid that interacts via a short range exponential attraction, the model parameter $\epsilon$ is identified with $\frac{3}{2}\frac{c_p}{c_p^\ast}\frac{R}{R_g}$ the values of which are experimentally known. Figure~\ref{fig:GPrime}(d) shows that the hybrid-PDT theory qualitatively captures all the features of the non-monotonic elastic response observed in the experiments. However, quantitative deviations are evident, which seem inevitable for multiple distinct reasons. These include that the theory is approximate, the model is not a polymer-colloid mixture, and the functional form of the effective attraction is not an exact representation of the depletion attraction.  In addition,  the experimental data of ref.~\cite{Pham2008} is based on measurement of the finite-frequency elastic modulus, which is not of a flat frequency-independent form for all the samples. Hence, there is some polymer-concentration-dependent uncertainty in extracting a single value of the plateau elastic modulus from frequency-dependent measurements, as discussed by the authors of ref.~\cite{Pham2008}.  \\

\section{INTERMEDIATE TIME AND LENGTH DYNAMICS}\label{Section4}
\subsection{A.	Dynamic Free Energy, Length and Energy Scales}
As the attractive interaction is increased, the dynamic behavior of dense sticky particle fluids undergoes changes beyond the typical glass melting phenomenon discussed in the context of long time relaxation and flow. For attractive glasses, simulations have found that the mean squared displacement (MSD) exhibits a long sub-diffusive regime on length scales smaller than the particle diameter before the eventual crossover to diffusive Fickian motion \cite{Zaccarelli2009,Fullerton2020}. This suggests the dynamical importance of additional local, in-cage, time and length scales beyond the small transient localization length, but before the structural relaxation event associated with activated barrier crossing. Such effects can be potentially captured by the NLE theory spatially resolved dynamic free energy concept which predicts a displacement-dependent effective force on a particle driven by uphill thermal fluctuations.\\
\begin{figure}
    \centering
    \includegraphics[width=0.45\textwidth]{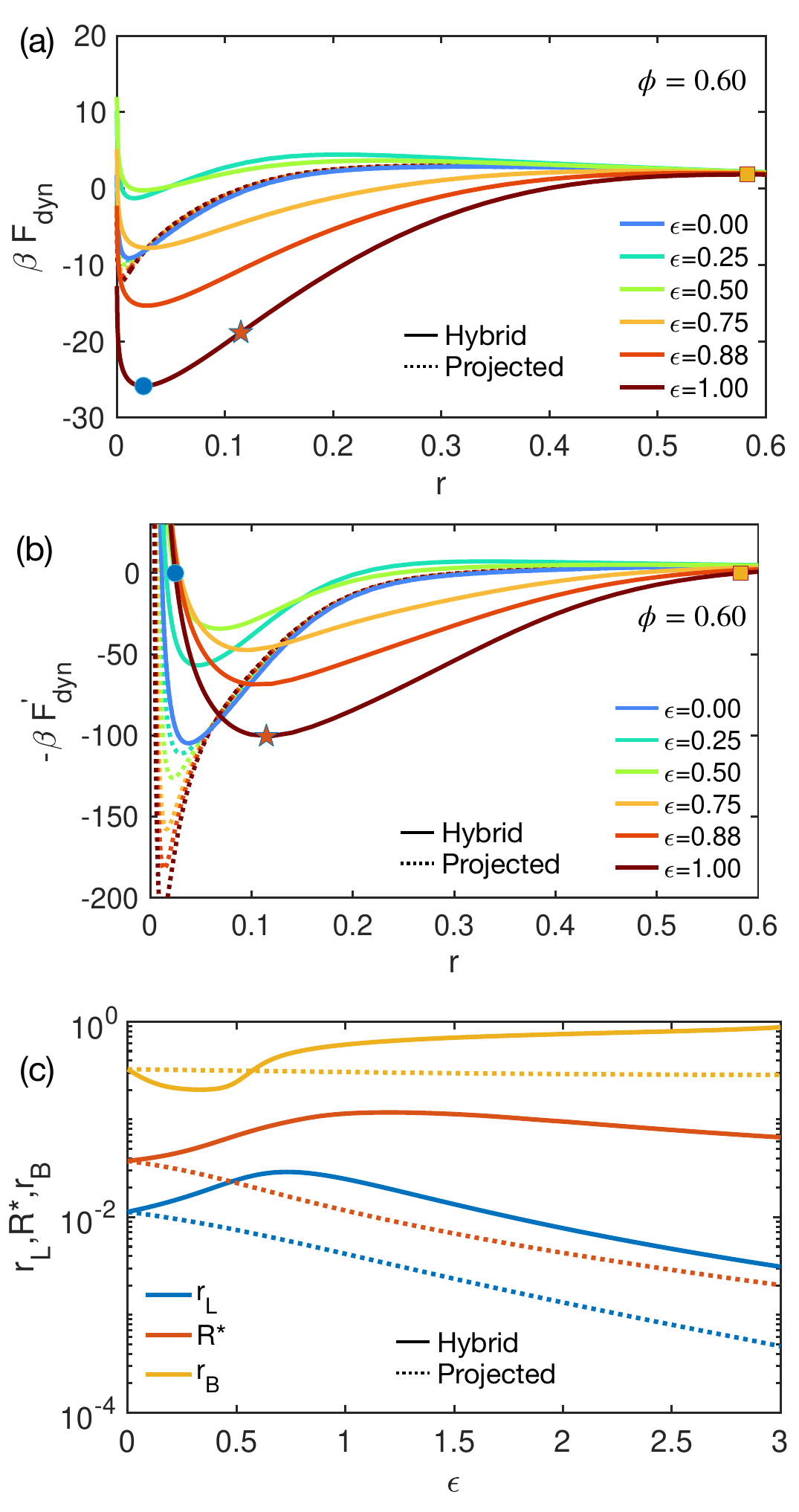}
    \caption{(a) Examples of the dynamic free energy for different attraction strengths and range $a=0.02$ at a packing fraction $\phi=0.60$. The solid curves are the calculations with the hybrid PDT vertex, while dotted lines are the corresponding projected ones. {\color{black}The dynamic free energy minimum for $\epsilon=1.0$ is indicated by the blue circle, the inflection point is marked by the red star, and the barrier location is denoted by the yellow square.} (b) The negative of first derivative of $F_{dyn}$ defines the effective force on a tagged particle. {\color{black}The indicated points are in the same position as those shown in the panel (a) above.} (c) Variation of different length scales with $\epsilon$.  Here, $r_L$ is the dynamic localization length, $R^\ast$ is the location of inflection point or the point of maximum force on the dynamic free energy, and $r_B$ is the barrier location.}
    \label{fig:FDyn}
\end{figure}

Figure~\ref{fig:FDyn}(a) shows example dynamic free energy curves for different attraction strengths of a short range $a=0.02$ system at a high $\phi=0.60$. The solid curves employ the hybrid vertex, while the dotted curves are for the projected theory. Figure~\ref{fig:FDyn}(b) shows the corresponding effective forces, $-\partial F_{dyn}/\partial r$. The projected dynamic free energy for the attractive systems resembles that of hard spheres but with a deeper well, indicating large confining forces. The absence of non-monotonic behavior with attraction strength in the localization length, local cage barrier, and all features of the dynamic free energy is evident. Conversely, the $F_{dyn}$ obtained using the hybrid dynamic vertex exhibit qualitative differences. Clear non-monotonic trends are seen such as the inflection point of maximum confining force shifting outward with increasing attraction strength (see Fig.~\ref{fig:FDyn}(c)), and the force magnitude does not change much despite the local barrier becoming much larger. These features result in a wide window of displacements between the localization length and barrier locations, potentially providing a theoretical basis for the long sub-diffusive regime of the MSD observed in simulations \cite{Zaccarelli2009,Fullerton2020}. \\

Figure~\ref{fig:FDyn}(c) summarizes the variation of the different dynamical length scales: localization length ($r_L$), inflection point ($R^\ast$), and barrier location ($r_B$), as a function of attraction strength for the projected and hybrid force vertex based theories. In the projected scenario, the largest length scale (barrier location) remains invariant with respect to attraction strength. However, in the hybrid PDT theory, it exhibits a non-monotonic trend, consistent with dynamic glass melting. Additionally, the point of maximum force shifts outward with increasing attraction strength in the hybrid theory, while it moves inward in the projected theory. These displacement dependent features must have major consequences for the stochastic trajectories of a moving particle predicted by the NLE evolution equation, and hence various ensemble averaged properties which weight the trajectories in different ways. \\

\subsection{Stochastic Trajectories}
Within NLE theory, one can obtain stochastic trajectories by numerically solving the overdamped non-linear Langevin equation (Eq.~(\ref{eqn:NLE})) using the dynamic free energy (Eq.~(\ref{eqn:Fdyn})). This method was successfully applied by Saltzman and Schweizer. \cite{Saltzman2006,Saltzman2006_2} to study the MSD and non-Gaussian or single particle dynamic heterogeneity (DH) effects (and many other properties) associated with stochastic trajectory fluctuations in dense hard sphere fluids. Note these are entirely dynamic fluctuation effects since possible structural disorder would induce a distribution of dynamic free energies. This aspect has been studied within the NLE and ECNLE theory frameworks \cite{Schweizer2004,Xie2020}, but is not our present focus. Rather, we employ the same stochastic trajectory method as before \cite{Saltzman2006,Saltzman2006_2} but now solve the NLE  for sticky particles based on the hybrid-PDT approach. \\

We first briefly review the technical details as discussed in depth previously \cite{Saltzman2006,Saltzman2006_2}. For each trajectory, Eqs.~(\ref{eqn:NLE}), (\ref{eqn:HybridForceVertex}), and (\ref{eqn:Fdyn}) are numerically solved. We perform $10^5$ independent simulations that are initiated at $r=r_L$ and propagated until the particle escapes the cage defined as crossing the barrier at $r=r_B$. Subsequently, the particle is considered to follow simple Brownian dynamics based on a Langevin equation with an enhanced friction coefficient of $\zeta_s\rightarrow\zeta_s+\zeta_{hop}$. The hopping friction coefficient is computed based on an elementary Fick’s law procedure:
\begin{equation}\label{eqn:Xihop}
    \frac{1}{\zeta_{hop}}=\frac{1}{N}\sum_i\frac{r_B^2}{6t_i^{FPT}} .
\end{equation}
Here, ${t_i}^{FPT}$ represents the first passage time of a specific trajectory $i$ to cross the barrier. We use a total of $50,000$ samples to obtain $\zeta_{hop}$. This is the identical algorithm as used in Ref.\cite{Saltzman2006,Saltzman2006_2}. Our focus here is the ensemble-averaged MSD and non-Gaussian parameter. In this section we are only interested in the relatively small displacements that define the “in cage” intermediate regime. Hence, collective elastic effects, which scale as the $4^{th}$ power of displacement, are unimportant.\\
\subsection{Mean Square Displacement and NonGaussian Parameter}
\begin{figure}
    \includegraphics[width=0.47\textwidth]{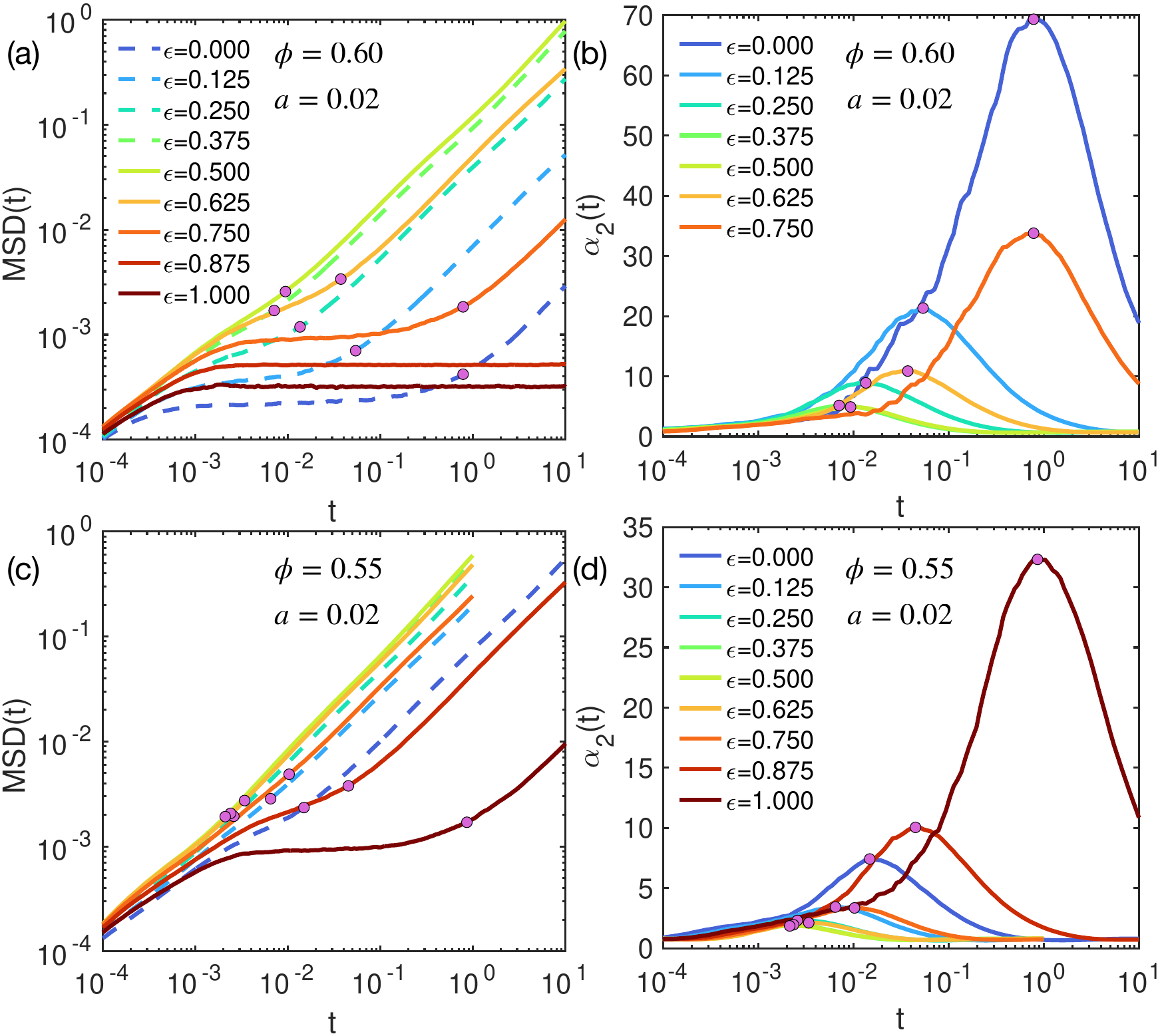}
    \caption{(a) and (c) Mean Squared Displacement (MSD) {\color{black} based on the hybrid PDT vertex} as a function of dimensionless time corresponding to the high packing fraction states of $\phi=0.60$ and $\phi=0.55$, respectively, for various indicated values of attraction strength. The attraction range is fixed at $a = 0.02$. The solid circles indicate the time and MSD value corresponding to the most non-Fickian or sub-diffusive state defined as when the local time scaling power law exponent is a minimum. The corresponding Non-Gaussian parameters (NGP) are shown in panels (b) and (d), respectively. For visual clarity, the solid circles indicate the location of the maximum of the NGP. The color codes in the bottom panels are identical to those defined in the upper panels. 
}
    \label{fig:MSDNGP}
\end{figure}

Figures~\ref{fig:MSDNGP}(a) and (c) shows the MSD, $\left\langle\Delta r\left(t\right)^2\right\rangle=\left\langle\left|r\left(t\right)-r\left(0\right)\right|^2\right\rangle$, for different attraction strengths at a fixed short range and two high packing fractions of $\phi=0.60$ and $\phi=0.55$. The $\epsilon=0$ curve (blue) is the hard-sphere result which shows a significant sub-diffusive regime (plateau-like behavior) for both packing fractions. The glass melting behavior is manifest by an increased localization length (MSD plateau) and decrease of the early stage of the cage escape time. As attraction strength sufficiently grows, a reversal in behavior is predicted indicating the formation of physical bonds that slow down particle motion, in a manner that depends on the packing fraction.\\

Figure~\ref{fig:MSDNGP}(b) and (d) show the corresponding non-Gaussian parameters (NGP), $\alpha_2\left(t\right)=\frac{3}{5}\frac{\left\langle\Delta r^4\left(t\right)\right\rangle}{\left\langle\Delta r^2\left(t\right)\right\rangle^2}-1.0$ for different attraction strengths at $\phi=0.60$ and $\phi=0.55$, respectively. As expected, the NGP peak exhibits a non-monotonic behavior with increasing attraction strength, initially decreasing, and then steeply increasing. This indicates a large amount of trajectory heterogeneity on intermediate time and length scale in attractive glasses. \\

\begin{figure}
    \includegraphics[width=0.47\textwidth]{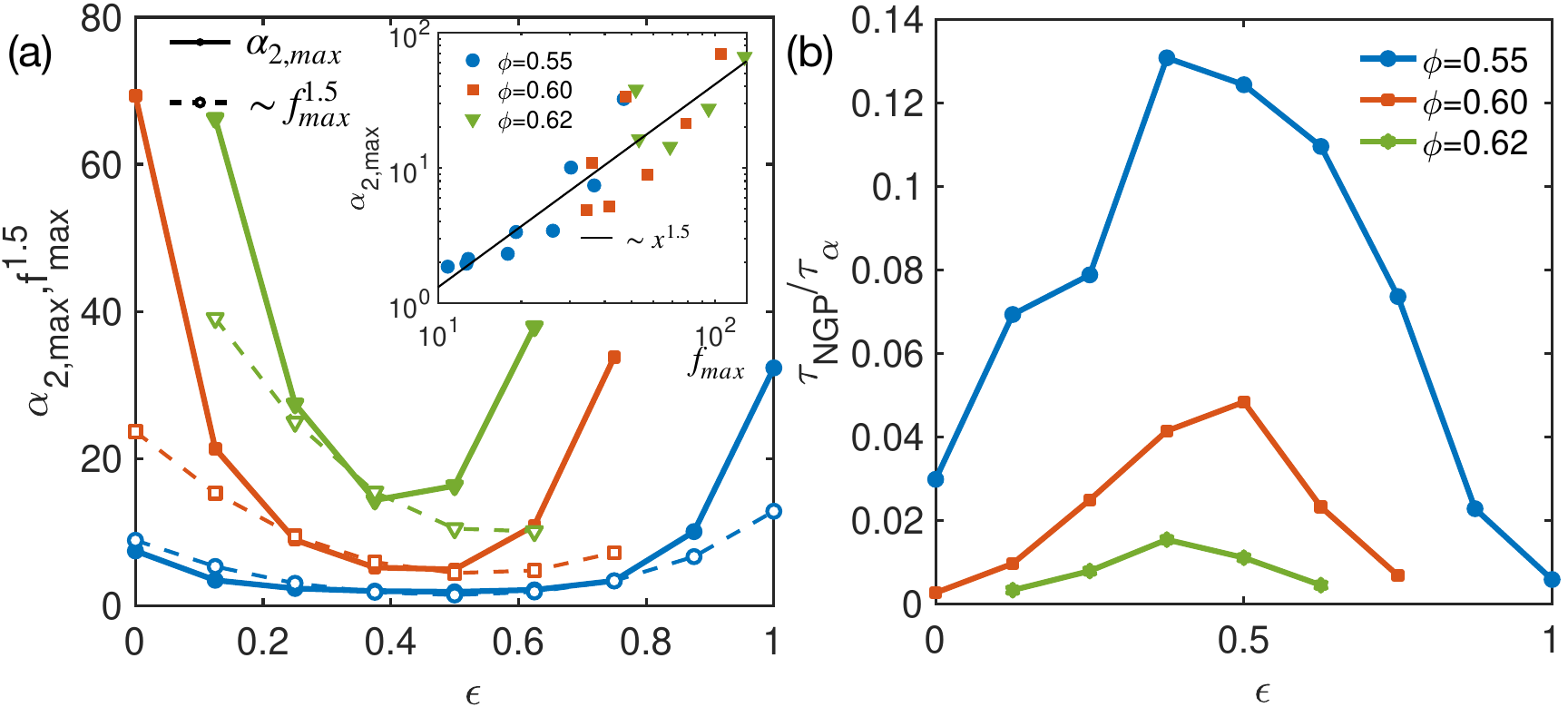}
    \caption{(a) The peak height of the non-Gaussian parameter ($\alpha_{2,max}$) for fluids with an attraction range of $a=0.02$ is shown (solid lines) to exhibit nonmonotonic behavior with attraction strength for different packing fractions. The inset presents a log-log cross plot of $\alpha_{2,max}$ versus the maximum confining force amplitude for three different high packing fractions. A global power law fit yields the relation $\alpha_{2,max}\propto f_{max}^{1.5}$. The main panel compares the attraction strength dependence of $f_{max}^{1.5}$ (dashed lines) and $\alpha_{2,max}$    (b) The corresponding timescale of the NGP peak scaled by the mean alpha time as a function of attraction strength for $3$ high packing fractions is depicted. }
    \label{fig:NGPMax}
\end{figure}
The evolution of the NGP peak value is shown in Fig.~\ref{fig:NGPMax}(a), and the timescale of the maximum scaled by the respective mean alpha time in Fig.~\ref{fig:NGPMax}(b). The NGP peak is known to occur on a timescale smaller than the alpha relaxation times, and increasingly so as the alpha time grows. This behavior is observed for the three packing fractions shown, and is correlated with the NGP peak height. The simplest physical intuition for the location in time and displacement, and peak amplitude of the NGP, is it qualitatively correlates with the theoretically predicted maximum effective caging or confining force $f_{max}=-\frac{\partial F_{dyn}}{\partial r}|_{r=R^\ast}$ \cite{Saltzman2006}. The inset of Fig.~\ref{fig:NGPMax}(a) shows a log-log cross plot between these two quantities, motivated by prior the NLE theory prediction for dense hard sphere fluids that $\alpha_{2,max}\propto f_{max}^{7/4}$ \cite{Saltzman2006}. Since the dense attractive fluids of present interest involve three variables (packing fraction, the attraction strength, and range) versus only one for hard spheres, we do not expect any collapse of the data. The results in the inset of Fig.~\ref{fig:NGPMax}(a) confirm this expectation, although a global fit does yield a rough power law with an exponent of $1.5$, close to the $1.75$ value found for hard spheres \cite{Saltzman2006}. {\color{black} The difference between effective exponents of $1.50$ and $1.75$ is not surprising given the presence of physical bonding in addition to caging changes the precise form of the dynamic free energy as a function of displacement, and hence the effective spatially-resolved forces. However, interpretation of this modest quantitative difference of apparent exponent is difficult.} The main panel of Fig.~\ref{fig:NGPMax}(a) plots $f_{max}^{1.5}$ and the peak NGP as a function of attraction strength. A strong non-monotonic evolution with attraction strength is found for the maximum NGP, that grows in amplitude as packing fraction increases. Crudely, this behavior is captured by the evolution of the maximum caging force, thereby providing a simple zeroth order interpretation.\\

\subsection{MSDs and NGPs Along an Isochrone}
\begin{figure}
    \includegraphics[width=0.47\textwidth]{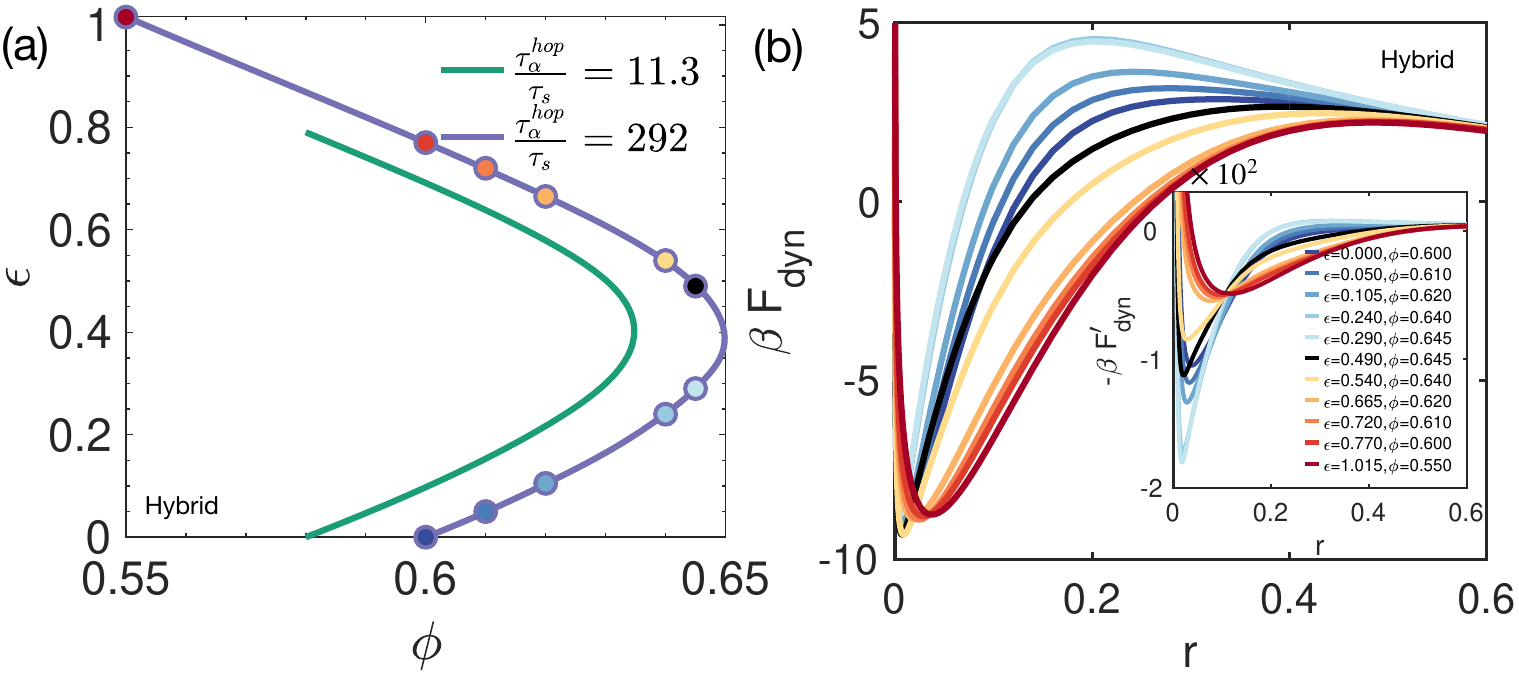}
    \caption{(a) Two different isochrones {\color{black} based on using the hybrid-PDT vertex and} constructed based on a fixed value of the mean alpha hopping times corresponding to states of variable packing fraction and attraction strength at a common attraction range of $0.02$. The indicated dimensionless mean hopping times are $11.3$ and $292$, which correspond to local cage barriers of $8.4$ and $12.0$, respectively. The points indicate the specific isochronal state points studied in detail for $\tau_\alpha^{hop}=292$.  (b) The dynamic free energy curves for the state points indicated in panel (a), with the corresponding forces shown in the inset.  }
    \label{fig:MSDIsochrone}
\end{figure}
As discussed in the previous section, the variation of the MSDs and NGPs with increasing attraction strength at a fixed attraction range and packing fraction reflect the glass melting phenomenon on relatively short (per the MSD plateau) and intermediate (per the NGP peak magnitude and location) time and length scales; the corresponding behavior of the activated barrier hopping time ($\tau_\alpha$) is discussed in next section. Here, we seek to probe the changes of these short and intermediate length properties along an isochrone defined as a fixed mean activated barrier hopping time (longest relaxation time) with variable attraction strength and packing fraction.\\
\begin{figure*}
    \centering
    \includegraphics[width=0.95\textwidth]{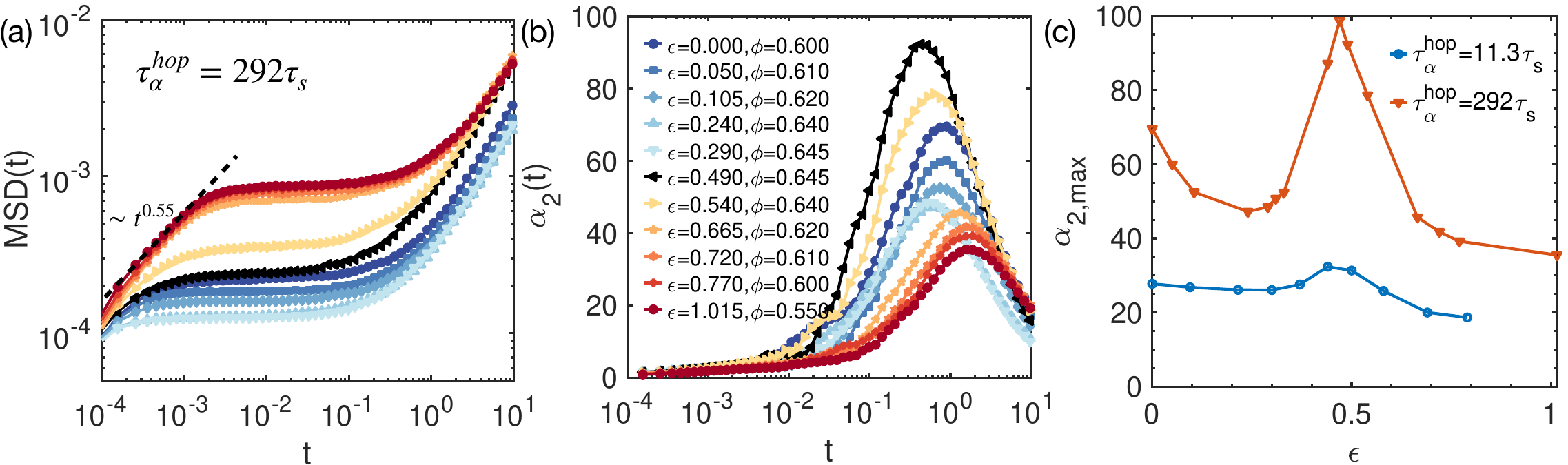}
    \caption{ {\color{black}Hybrid-PDT vertex based calculations of the} MSD (a) and NGP (b) along an isochrone (blue curve in Fig.~\ref{fig:MSDIsochrone}(a)) with fixed mean dimensionless hopping time, $\tau_\alpha^{hop}=292\tau_s$ corresponding to a local cage barrier of $12.0$ for an attraction range of $0.02$. Panel (c) plots the maximum of the NGP ($\alpha_{2,max}$) extracted from panel (b) as a function of attraction strength along two isochrones with the indicated mean hopping times differing by nearly a factor of 30. The corresponding MSD and NGP results for the $\tau_\alpha^{hop}=10\tau_s$ isochrone are shown in the SI.   
    }
    \label{fig:MSDIsochrone2}
\end{figure*}

Fig.~\ref{fig:MSDIsochrone}(a) shows two such isochrones with fixed mean alpha times that differ by a factor of $\sim30$. The isochronal curve corresponds to a non-monotonic evolution of packing fraction with increasing attraction strength, with one end of the curve corresponding to the hard sphere repulsive glass, and the other end an attractive glass at large $\epsilon$. Fig.~\ref{fig:MSDIsochrone}(b) shows the corresponding dynamic free energy curves along an isochrone when the pure hard sphere repulsive glass state occurs at $\phi=0.60$, while the inset presents the corresponding forces. The barrier heights are very similar which is an expected consequence of the isochronal alpha time condition. On the other hand, the evolution of the functional form of the dynamic free energy on smaller length scales is highly variable. Specifically, the localization length, barrier location, and displacement and magnitude of the maximum confining force state exhibit non-monotonic variations along the isochrone. The state of maximum confining force is located just below the nose feature of an isochrone. The localization length and barrier location shift to smaller displacements with increasing $\epsilon$ until the nose is reached at $\epsilon\sim0.4$, after which they reverse their dependence on attraction strength and move outwards. The non-monotonic evolution of the localization length can be understood as primarily a consequence of its non-monotonic packing fraction variation along an isochrone. In contrast, the non-monotonic evolution of the larger length scale barrier location is more subtle since it cannot be explained by the non-monotonic variation of packing fraction. The maximum confining force initially increases with attraction strength, and then decreases to significantly lower values beyond the nose. This reduction in maximum confining force, and its outwards shift in particle displacement in the attractive glass (high $\epsilon$) regime, is expected to result in a broader sub-diffusive regime of the MSD. The latter effect has been observed in simulations \cite{Fullerton2020,Zaccarelli2009}. \\

Fig.~\ref{fig:MSDIsochrone2}(a) presents the MSD plots for the state points indicated in Fig.~\ref{fig:MSDIsochrone}(a). Their overall form is, to leading order, qualitatively similar, with non-monotonic features clearly evident. The intermediate time plateau (dynamic localization length) first modestly decreases with increasing attraction strength along the isochrone as a consequence of the increasing packing fraction. Based on this behavior, one can say the introduction of weak attractions \textit{along} an isochrone results in more localized repulsive-like glasses. As the attraction strength increases further, the corresponding packing fraction along the isochrone then decreases, resulting in a MSD plateau that significantly increases, corresponding to weaker localization. This behavior does tends to saturate at high enough attraction strength. These trends are all qualitatively consistent with what is expected based on the rich evolution of the length-scale-dependent dynamic free energy in Fig.~\ref{fig:MSDIsochrone}(b).  Moreover, for the attractive glass cases, an interesting apparent power law sub-diffusive regime emerges on time scales before the plateau is reached. Though differing in detail given differences in the models studied (form of the potential, its range, polydispersity) and state points, this behavior seems roughly akin to what has been observed in simulations \cite{Fullerton2020,Zaccarelli2009} of attractive glasses, and which has been suggested to be associated with a broad range of dynamically-relevant length scales. In the context of the present microscopic theory, the latter is naturally captured via the predicted distinctive changes of the shape of the dynamic free energy curves in Fig.~\ref{fig:MSDIsochrone}(b), and hence displacement-dependent force on a moving tagged particle, which undergo nonperturbative changes as attractions become strong enough. {\color{black}Specifically, we beleive this feature relates to the fact that in the attractive glass regime the dynamic free energy is predicted to be less steep for displacements far below the barrier (see Fig.~\ref{fig:MSDIsochrone}(b)).}\\

Fig.~\ref{fig:MSDIsochrone2}(b) shows the temporal evolution of the NGP for the same systems as in Fig.~\ref{fig:MSDIsochrone2}(a), while Fig.~\ref{fig:MSDIsochrone2}(c) plots the peak value of the NGP, both along the isochrone with increasing $\epsilon$. Starting from the hard sphere case, the value of the maximum NGP decreases along an isochrone, indicating a reduction of single particle dynamic heterogeneity effects associated with stochastic trajectory fluctuations. As attraction strength further increases and the nose of an isochrone is approached ($\epsilon=0.4$), the NGP maximum amplitude sharply increases, resulting in an overall non-monotonic behavior of this property. Upon further increase of $\epsilon$ into the attractive glass regime, the NGP amplitude goes through a maximum and again decreases. Overall, a rich doubly non-monotonic evolution of this property along an isochrone is predicted. Physically, as one moves along an isochrone starting from the pure hard sphere system, there is a transition from caging-dominant repulsive glass behavior to bonding-dominant attractive glass states. {\color{black} The predicted peak in the maximum NGP at an intermediate value of attraction strength reflects a frustration-like effect on particle trajectories. Specifically via the spatially-resolved bonding and caging kinetic constraints as embedded in the dynamic free energy as the attraction strength and packing fraction are simultaneously varied along an isochrone at a fixed barrier or mean hopping time.} \\

The corresponding MSD and NGP results along an isochrone for the smaller mean hopping time system of $\tau_\alpha^{hop}=11.3$ (green curve in Fig.~\ref{fig:MSDIsochrone}(a)) are shown in the Supplementary Information. Fig.~\ref{fig:MSDIsochrone2}(c) shows the corresponding behavior of the NGP maximum amplitude. Beyond the expected smaller overall magnitude of the NGP maximum for these less localized states, the functional dependence on attraction strength is qualitatively the same, albeit the amplitude of the non-monotonic features is much weaker. \\

\section{LONG TIME DYNAMICS AND KINETIC ARREST PHASE DIAGRAMS }\label{Section5}
We now consider the long time relaxation that is controlled mainly by the height of the local cage and the longer range collective elastic barriers, and the corresponding activated mean alpha relaxation times and isochronal kinetic arrest maps.  \\

\subsection{Local Cage and Collective Elastic Barriers}
\begin{figure}
    \centering
    \includegraphics[width=0.45\textwidth]{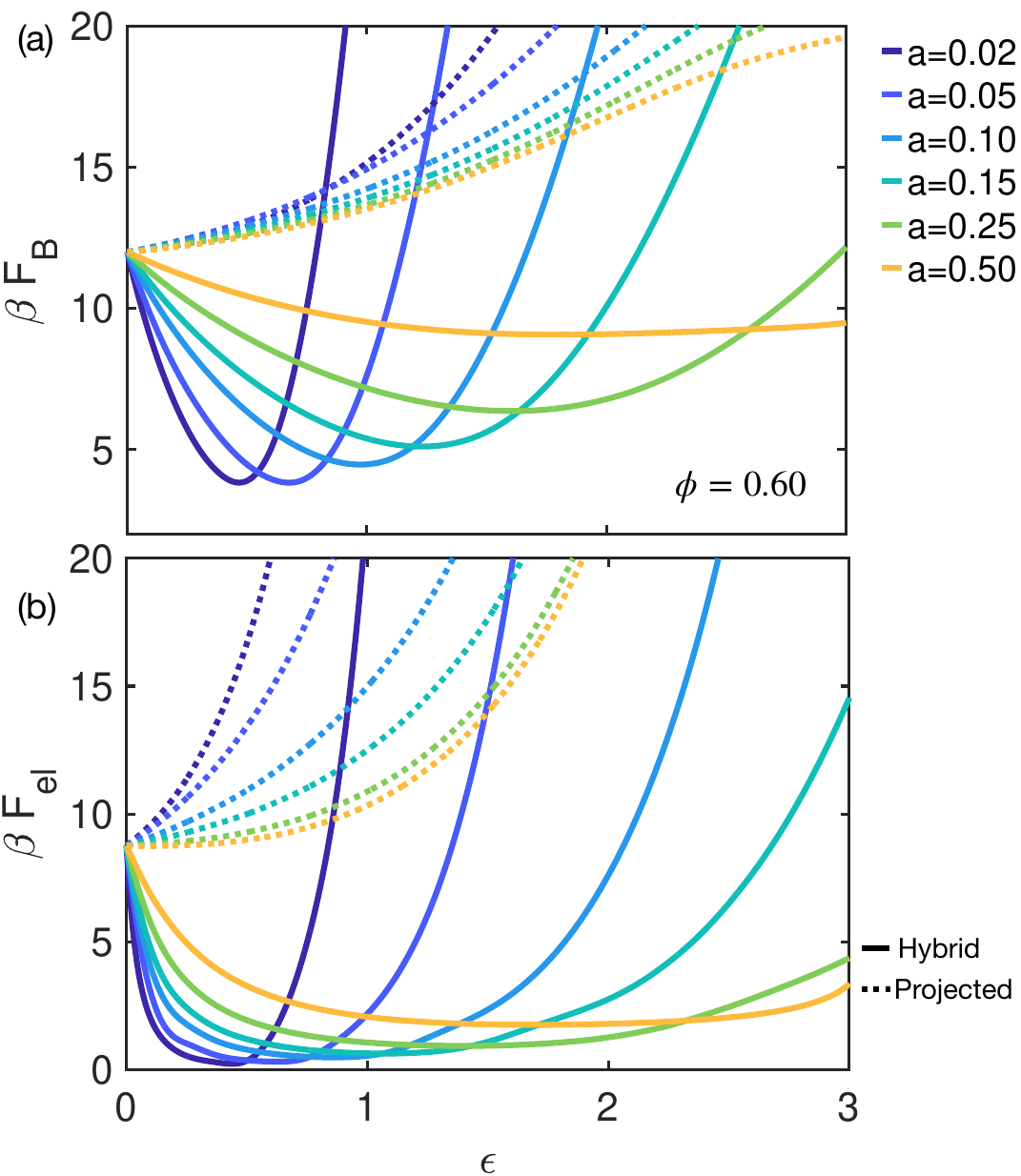}
    \caption{(a) Local barrier in units of the thermal energy as a function of attraction strength for different attraction ranges at a fixed value of packing fraction $\phi=0.60$. (b) Corresponding results for the collective elastic barrier. {\color{black} All solid curves employ the hybrid-PDT vertex while dotted curves employ the projected analog.}}
    \label{fig:Fb}
\end{figure}

Figure~\ref{fig:Fb} shows the evolution of the local cage barrier as a function of attraction strength for various ranges at a high packing fraction of $\phi=0.60$. Importantly, this state is very far beyond the ideal NMCT boundary (a dynamic crossover), and the local barrier does not exhibit any non-monotonic behavior in the projected description (dotted curves), whereas strong non-monotonic upturns are evident with the hybrid dynamic vertex (solid curves). Notably, the barrier height undergoes a substantial increase after the upturn, indicating strengthening of physical bonds. The non-monotonic feature is also predicted to shift to larger attractive strengths with increasing range, accompanied by a reduction in magnitude. The maximum glass melting occurs at $\epsilon=0.5$ for a range of $a=0.02$, a value that is roughly two times smaller than found for the localization length at $\phi=0.60$ (see Fig.~\ref{fig:RLoc}). More generally, the minimum local barrier state occurs at larger attraction strength with increasing range, and also is not as deeply suppressed. When the attraction range reaches the large value of $a=0.5$ typical of nonpolar atoms and molecules (like the LJ potential) with slowly varying attractions, the barrier melting effect has essentially disappeared, as physically expected.\\

\begin{figure}
    \centering
    \includegraphics[width=0.45\textwidth]{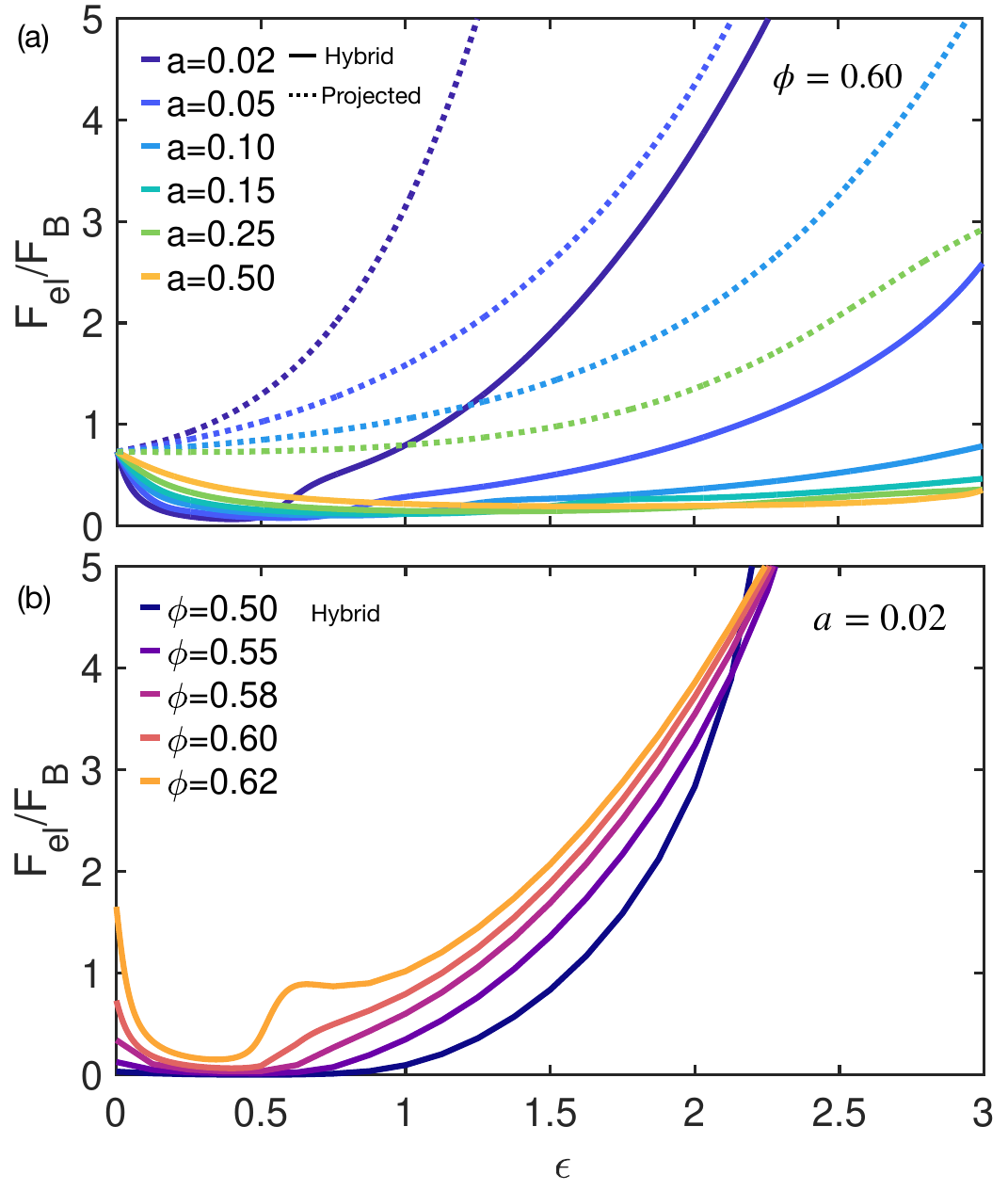}
    \caption{(a)The ratio of elastic barrier to the local barrier with increasing attraction strength for different attraction ranges for a packing fraction $\phi=0.60$. (b) The same for different packing fractions for range of $a=0.02$. {\color{black} All solid curves employ the hybrid-PDT vertex while dotted curves employ the projected analog.}}
    \label{fig:FElByFb}
\end{figure}

Fig.~\ref{fig:Fb}(b) shows the corresponding results for the collective elastic barrier as a function of attraction strength for various ranges at the same high packing fraction in the attractive glass regime. One sees that based on the hybrid dynamic vertex, the elastic barrier first decreases rapidly, and remains small until the local barrier starts to increase. Eventually, the elastic barrier overtakes the local barrier in a manner similar to the behaviour observed with increasing packing fraction. In qualitative contrast, for the corresponding projected case, as can be inferred from the increased curvature at $F_{dyn}$ minima in Fig.~\ref{fig:FDyn}, the elastic barrier increases monotonically with attraction strength.\\

The ratio of the elastic to local cage barriers is the key measure of “cooperativity” and dynamic fragility in ECNLE theory \cite{Xie2016,Mei2020}. Figure~\ref{fig:FElByFb}(a) shows a nonmonotonic behaviour is predicted as a function of attraction strength, with a low amplitude, nearly flat “glass melting” intermediate regime. This strong suppression occurs for almost all spatial ranges and extends to rather high values of attraction in the hybrid theory. Thus, for attraction ranges of $a = 0.1$ and higher, the alpha process becomes less cooperative or less fragile in this sense. However, for very small attraction ranges, the barrier ratio does become significantly larger than unity at high enough attraction. Overall, as attraction strength grows from zero the system evolves from a fragile repulsive fluid, to an effectively strong fluid, and then again a fragile system at high enough attraction strength. Fig.~\ref{fig:FElByFb} shows that the opposite behaviors are predicted based on the projected force vertex where the barrier ratio always increases strongly with attraction strength.\\

Figure~\ref{fig:FElByFb}(b) presents calculations of the packing fraction variation of the barrier ratio. For a low enough packing fraction, the degree of cooperativity remains small with increasing attraction strength, showing very little non-monotonicity. However, at sufficiently high attraction, physical bonding impacts relaxation, marking the onset of an attractive glass. Since generally increasing packing fraction increases the relative importance of collective elasticity, the ratio is larger, for almost all attractions strengths, as the fluid packing fraction increases. Subtle curve crossings emerge at very high attractions.\\
\subsection{Mean Alpha Relaxation Times and Isochronal Kinetic Arrest Boundaries}
\begin{figure}
    \centering
    \includegraphics[width=0.40\textwidth]{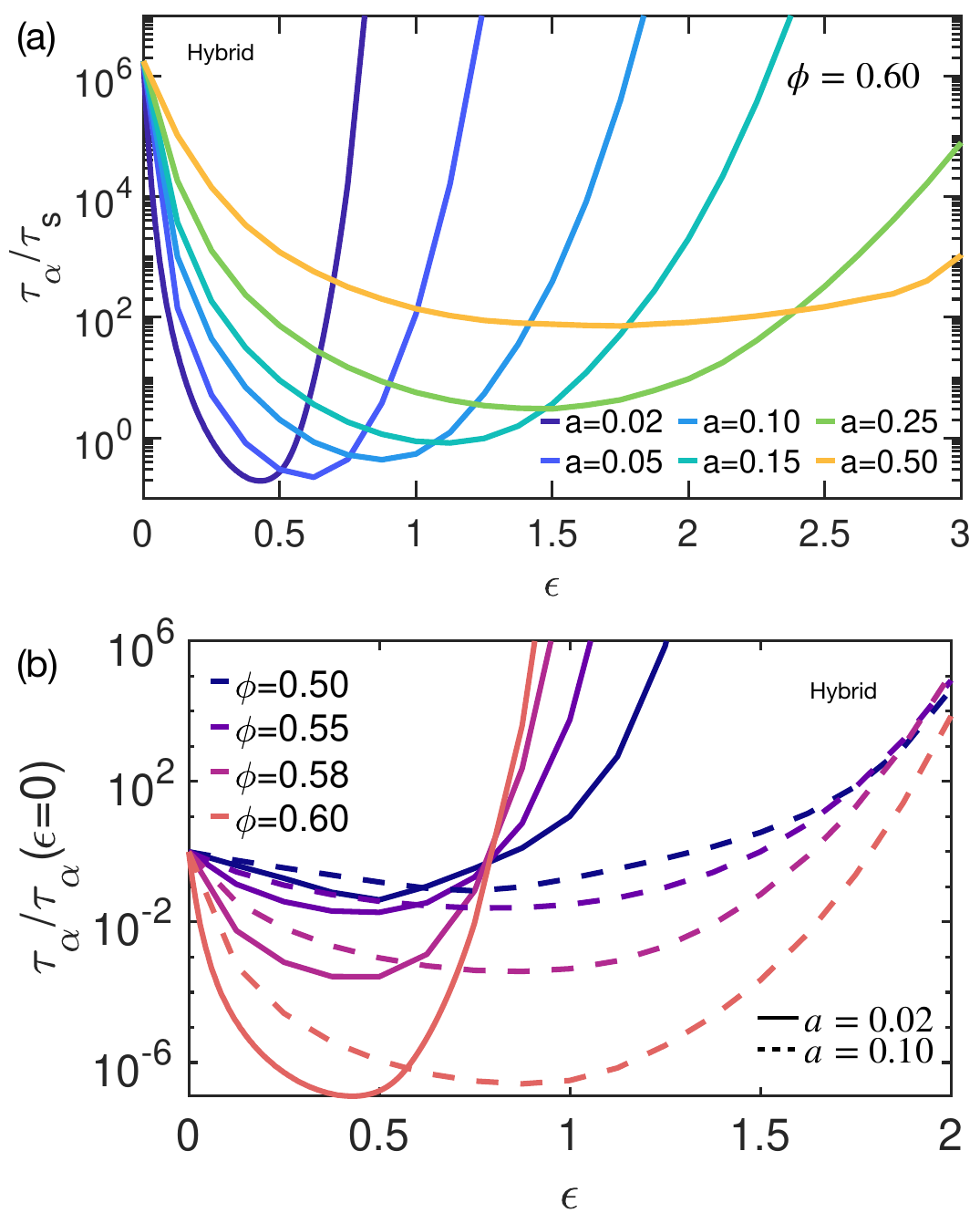}
    \caption{(a) Alpha time (in units of the short process timescale, $\tau_s$) as a function of attraction strength for different attraction ranges at a fixed value of packing fraction $\phi=0.60$. (b) Mean alpha time now in units of its analog for the hard sphere fluid as a function of attraction strength for different packing fractions. Solid lines are for attraction range $a=0.02$, while the dashed lines are for $a=0.10$. {\color{black}All curves employ the hybrid-PDT vertex.}}
    \label{fig:TauAlpha}
\end{figure}
We now employ ECNLE theory and Eq.~(\ref{eqn:TauAlpha}, \ref{eqn:TauAlphaEl})  to study the alpha relaxation times, and the corresponding kinetic arrest boundaries based on a chosen timescale criterion.  An example of such isochrone boundaries has been briefly discussed theoretically previously \cite{Ghosh2019} (and also in simulation \cite{Fullerton2020}) and shown to exhibit glass melting features at extremely large packing fractions, and here are studied in much more depth. \\

Figure~\ref{fig:TauAlpha}(a) shows the non-monotonic evolution of alpha times as a function of attraction strength for different spatial ranges at a fixed high packing fraction based on the hybrid PDT ECNLE approach. The trends closely align with the predictions for total barriers, as expected. The maximum glass melting effect of  $\sim 7$ decades of speed up is observed for the shortest range attraction, consistent with the predicted $14$ units decrease of the total barrier. On the other hand, when comparing the maximum speed-up achieved with the introduction of an attractive potential at different packing fractions, the speed-up increases at higher $\phi$, as illustrated in Fig.~\ref{fig:TauAlpha}(b). This indicates that the slower the baseline repulsive system is,  the greater its accelerated relaxation is at modest attraction strengths.   \\

\begin{figure}
    \includegraphics[width=0.45\textwidth]{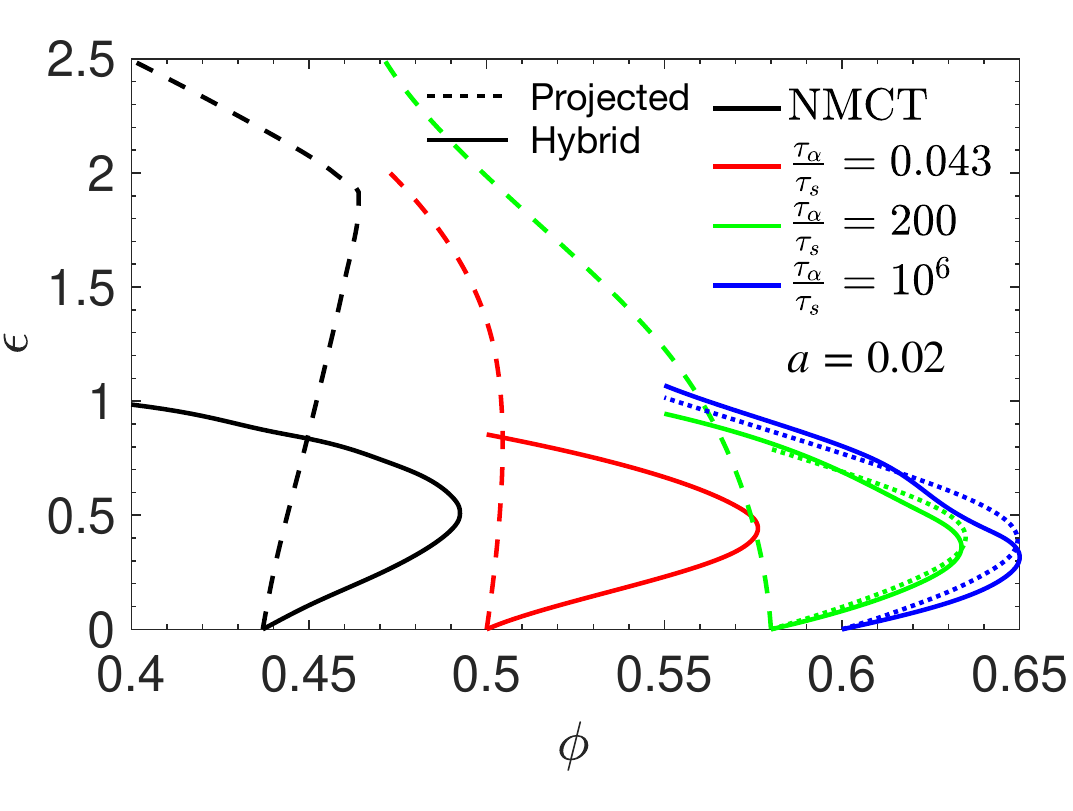}
    \caption{Kinetic arrest map is depicted as isochrones for different reduced mean alpha relaxation time criteria, constructed using projected (dashed) and hybrid (solid) theory. The black lines represent NMCT phase boundaries for each case. Solid lines correspond to $\tau_\alpha/\tau_s$ including the elastic barrier contributions, while the dotted lines represent isochrones obtained by considering only local cage barriers.}
    \label{fig:isochrones}
\end{figure}

Calculations of isochronal maps, curves of constant reduced alpha time in $\phi-\epsilon$  space, are shown in Fig.~\ref{fig:isochrones}. Their basic forms are relatively similar with increasing $\tau_{\alpha}$ magnitude kinetic arrest criterion. Note that Fig.~\ref{fig:isochrones} extends the idealized NMCT kinetic arrest boundary (black curves) to much larger packing fractions of typical experimental relevance. The dashed curves represent the isochrones/kinetic arrest boundaries based on the projected force vertex, while the solid curves depict those with the hybrid-PDT vertex. Each color corresponds to different choices of the alpha relaxation time that defines the kinetic arrest criterion. This plot clearly demonstrates that NMCT theory with the projected vertex fails to predict any glass melting at such large packing fractions. While arbitrarily shifting ideal MCT theory curves to larger $\phi$ might seem to indicate this behaviour, we believe this is not justifiable since the physics at high concentrations involves activated dynamics and large, but finite, relaxation times. On the other hand, the hybrid-PDT ECNLE theory does capture the non-monotonic behavior in Fig.~\ref{fig:isochrones} as a consequence of activated relaxation and explicit treatment of attractive forces. We note that the recent GMCT work \cite{Luo2021} does predict a significant shift of the ideal kinetic arrest boundaries to higher packing fractions. However, the underlying theory remains MCT-like in that it adopts the vertex projection strategy, and predicts the alpha relaxation time diverges and other dynamical singularity features remain. The latter is in qualitative contrast to the central importance of activated dynamics and explicit treatment of attractive forces in the present ECNLE theory.\\

The dotted curves in Fig.~\ref{fig:isochrones} show the corresponding hybrid vertex ECNLE theory isochrones obtained by considering only the local barriers. As seen in Figs.~\ref{fig:Fb} and \ref{fig:FElByFb}, the elastic barriers diminish rapidly with increasing attraction strength, and to a greater extent with increasing $\phi$. This shifts the isochrones based on only local cage barriers to higher attraction strength before the nose feature is reached. However, the elastic barrier grows rapidly after the glass melting point is reached, leading to a curve crossing of isochrones with and without elastic barriers. Overall, our results based only on local cage barriers are qualitatively the same in form as those that include the elastic barrier. This nontrivial result reflects the predicted strong correlation between elastic and local barriers \cite{Mirigian2014,Mirigian2013}.
\section{CONCLUSIONS AND FUTURE DIRECTIONS  }\label{Section6}
We have utilized the hybrid-PDT ECNLE theory to study, for the first time, the problem of elastic modulus re-entrancy in dense fluids composed of spheres interacting via short-range attractions at very high packing fractions far beyond the ideal MCT nonergodicity boundary. The dynamic force constraints (vertex) are formulated without adopting the literal projection approximation for constructing an effective attractive force from structural pair correlations. Instead, it retains an explicit treatment of the bare attractive forces that drive transient physical bond formation, while a projection approximation is employed for the singular hard-sphere potential. The resultant interference between repulsive and attractive forces contributions to the dynamic force vertex results in the prediction of localization length and elastic modulus re-entrancy, qualitatively consistent with experiments \cite{Pham2008}.  Such elastic re-entrancy is not captured by ideal MCT based on force projection, nor apparently its GMCT extensions \cite{Luo2021}. {\color{black} Of course, as we mentioned above,  in experiments the elastic re-entrancy effect can depend on measurement frequency. As a future direction, our approach can be extended to address this by explicitly analyzing the stress relaxation storage modulus in Fourier space.}\\

The non-monotonic evolution of the structural alpha relaxation time predicted by ECNLE theory with the hybrid PDT approach has been explored in far greater depth than previously \cite{Ghosh2019} as a function of packing fraction, attraction strength, attraction range, and under isochronal conditions. In addition, a detailed analysis of the length and energy scales of the dynamic free energy has been performed. Comparisons of these results with the corresponding predictions of ideal MCT based on projection and no activated dynamics, and also ECNLE and NLE theory based on projection, reveal large qualitative differences.\\

We have also investigated, for the first time, the consequences of stochastic trajectory fluctuations intrinsic to the NLE evolution equation description on intra-cage single particle dynamics with variable strength of attractions. The mean square displacement and non-Gaussian parameter were determined. The numerically observed \cite{Fullerton2020} large single particle dynamical heterogeneity effects for attractive glasses are captured as demonstrated by the rapidly increasing amplitude of the non-Gaussian parameter with packing fraction and a non-monotonic evolution with attraction strength. These behaviors reflect thermally driven “uphill” particle motion on the dynamic free energy profile, and have nothing to do with ideal MCT singularities which do not exist in NLE or ECNLE theory in the presence of thermal fluctuations. We have also constructed dynamic arrest boundaries based on activated relaxation determined  isochrones, which display the classic non-monotonic glass melting form. The latter behavior arises directly from the explicit treatment of attractive forces (hybrid PDT) and activated motion, and disappears if the standard full projection approximation is employed in ECNLE theory. These new results appear to be in qualitative accord with recent simulations that have employed swap-Monte Carlo \cite{Fullerton2020}. \\

Looking to the future, we suggest new simulations and experiments can more deeply test our results for both intermediate time and length scale single particle dynamics, and long time scale based isochronal kinetic arrest maps. Concerning new theory development, the present study, in combination with recent advances in the ECNLE theory of nonlinear rheology of dense glass forming hard sphere fluids and colloidal suspensions \cite{Ghosh2023_2,Ghosh2020_2}, sets the stage to construct a theory of nonlinear rheology of dense attractive colloidal suspensions. Due to the competition between caging and bonding, such systems are experimentally known \cite{Pham2008,Kaumakis2011,Moghimi2020} to exhibit the remarkable phenomenon (with nonuniversal features) of ``double yielding,'' for which no microscopic theory exists to date. Our work in this direction will be reported in a forthcoming publication. \\


\section*{Conflicts of interest}
There are no conflicts to declare.
\section*{Data availability statement}
The main text, or SI, contains all the data and the theoretical basis to construct the data.

\section*{Acknowledgements}
The authors acknowledge support from the Army Research Office via a MURI grant with Contract No. W911NF-21-0146.  We thank Ashesh Ghosh for helpful discussions. K.S.S. thanks Thomas Voightman for clarifying correspondence related to ideal MCT.



\balance


\bibliography{StickySphereEquil} 

\providecommand*{\mcitethebibliography}{\thebibliography}
\csname @ifundefined\endcsname{endmcitethebibliography}
{\let\endmcitethebibliography\endthebibliography}{}
\begin{mcitethebibliography}{79}
\providecommand*{\natexlab}[1]{#1}
\providecommand*{\mciteSetBstSublistMode}[1]{}
\providecommand*{\mciteSetBstMaxWidthForm}[2]{}
\providecommand*{\mciteBstWouldAddEndPuncttrue}
  {\def\EndOfBibitem{\unskip.}}
\providecommand*{\mciteBstWouldAddEndPunctfalse}
  {\let\EndOfBibitem\relax}
\providecommand*{\mciteSetBstMidEndSepPunct}[3]{}
\providecommand*{\mciteSetBstSublistLabelBeginEnd}[3]{}
\providecommand*{\EndOfBibitem}{}
\mciteSetBstSublistMode{f}
\mciteSetBstMaxWidthForm{subitem}
{(\emph{\alph{mcitesubitemcount}})}
\mciteSetBstSublistLabelBeginEnd{\mcitemaxwidthsubitemform\space}
{\relax}{\relax}

\bibitem[Angell \emph{et~al.}(2000)Angell, Ngai, McKenna, McMillan, and
  Martin]{Angell2000}
C.~A. Angell, K.~L. Ngai, G.~B. McKenna, P.~F. McMillan and S.~W. Martin,
  \emph{Journal of Applied Physics}, 2000, \textbf{88}, 3113–3157\relax
\mciteBstWouldAddEndPuncttrue
\mciteSetBstMidEndSepPunct{\mcitedefaultmidpunct}
{\mcitedefaultendpunct}{\mcitedefaultseppunct}\relax
\EndOfBibitem
\bibitem[Berthier and Biroli(2011)]{Berthier2011}
L.~Berthier and G.~Biroli, \emph{Reviews of Modern Physics}, 2011, \textbf{83},
  587–645\relax
\mciteBstWouldAddEndPuncttrue
\mciteSetBstMidEndSepPunct{\mcitedefaultmidpunct}
{\mcitedefaultendpunct}{\mcitedefaultseppunct}\relax
\EndOfBibitem
\bibitem[G{\"{o}}tze(2008)]{Gotze2008}
W.~G{\"{o}}tze, \emph{{Complex Dynamics od Glass-Forming Liquids: A
  Mode-Coupling Theory}}, Oxford University Press, 2008\relax
\mciteBstWouldAddEndPuncttrue
\mciteSetBstMidEndSepPunct{\mcitedefaultmidpunct}
{\mcitedefaultendpunct}{\mcitedefaultseppunct}\relax
\EndOfBibitem
\bibitem[Dhont(1996)]{dhont1996}
J.~Dhont, \emph{An Introduction to Dynamics of Colloids}, Elsevier Science,
  1996\relax
\mciteBstWouldAddEndPuncttrue
\mciteSetBstMidEndSepPunct{\mcitedefaultmidpunct}
{\mcitedefaultendpunct}{\mcitedefaultseppunct}\relax
\EndOfBibitem
\bibitem[Chaudhuri \emph{et~al.}(2007)Chaudhuri, Berthier, and
  Kob]{Chaudhuri2007}
P.~Chaudhuri, L.~Berthier and W.~Kob, \emph{Phys. Rev. Lett.}, 2007,
  \textbf{99}, 060604\relax
\mciteBstWouldAddEndPuncttrue
\mciteSetBstMidEndSepPunct{\mcitedefaultmidpunct}
{\mcitedefaultendpunct}{\mcitedefaultseppunct}\relax
\EndOfBibitem
\bibitem[Weeks \emph{et~al.}(2000)Weeks, Crocker, Levitt, Schofield, and
  Weitz]{Weeks2000}
E.~R. Weeks, J.~C. Crocker, A.~C. Levitt, A.~Schofield and D.~A. Weitz,
  \emph{Science}, 2000, \textbf{287}, 627--631\relax
\mciteBstWouldAddEndPuncttrue
\mciteSetBstMidEndSepPunct{\mcitedefaultmidpunct}
{\mcitedefaultendpunct}{\mcitedefaultseppunct}\relax
\EndOfBibitem
\bibitem[Miyagawa \emph{et~al.}(1988)Miyagawa, Hiwatari, Bernu, and
  Hansen]{Miyagawa1988}
H.~Miyagawa, Y.~Hiwatari, B.~Bernu and J.~P. Hansen, \emph{The Journal of
  Chemical Physics}, 1988, \textbf{88}, 3879–3886\relax
\mciteBstWouldAddEndPuncttrue
\mciteSetBstMidEndSepPunct{\mcitedefaultmidpunct}
{\mcitedefaultendpunct}{\mcitedefaultseppunct}\relax
\EndOfBibitem
\bibitem[Saltzman and Schweizer(2006)]{Saltzman2006}
E.~J. Saltzman and K.~S. Schweizer, \emph{Phys. Rev. E}, 2006, \textbf{74},
  061501\relax
\mciteBstWouldAddEndPuncttrue
\mciteSetBstMidEndSepPunct{\mcitedefaultmidpunct}
{\mcitedefaultendpunct}{\mcitedefaultseppunct}\relax
\EndOfBibitem
\bibitem[Saltzman and Schweizer(2006)]{Saltzman2006_2}
E.~J. Saltzman and K.~S. Schweizer, \emph{The Journal of Chemical Physics},
  2006, \textbf{125}, 044509\relax
\mciteBstWouldAddEndPuncttrue
\mciteSetBstMidEndSepPunct{\mcitedefaultmidpunct}
{\mcitedefaultendpunct}{\mcitedefaultseppunct}\relax
\EndOfBibitem
\bibitem[Hunter and Weeks(2012)]{Hunter2012}
G.~L. Hunter and E.~R. Weeks, \emph{Reports on Progress in Physics}, 2012,
  \textbf{75}, 066501\relax
\mciteBstWouldAddEndPuncttrue
\mciteSetBstMidEndSepPunct{\mcitedefaultmidpunct}
{\mcitedefaultendpunct}{\mcitedefaultseppunct}\relax
\EndOfBibitem
\bibitem[Bergenholtz and Fuchs(1999)]{Bergenholtz1999}
J.~Bergenholtz and M.~Fuchs, \emph{Physical Review E}, 1999, \textbf{59},
  5706–5715\relax
\mciteBstWouldAddEndPuncttrue
\mciteSetBstMidEndSepPunct{\mcitedefaultmidpunct}
{\mcitedefaultendpunct}{\mcitedefaultseppunct}\relax
\EndOfBibitem
\bibitem[Dawson \emph{et~al.}(2000)Dawson, Foffi, Fuchs, G\"otze, Sciortino,
  Sperl, Tartaglia, Voigtmann, and Zaccarelli]{Dawson2000}
K.~Dawson, G.~Foffi, M.~Fuchs, W.~G\"otze, F.~Sciortino, M.~Sperl,
  P.~Tartaglia, T.~Voigtmann and E.~Zaccarelli, \emph{Phys. Rev. E}, 2000,
  \textbf{63}, 011401\relax
\mciteBstWouldAddEndPuncttrue
\mciteSetBstMidEndSepPunct{\mcitedefaultmidpunct}
{\mcitedefaultendpunct}{\mcitedefaultseppunct}\relax
\EndOfBibitem
\bibitem[Zaccarelli \emph{et~al.}(2002)Zaccarelli, Foffi, Dawson, Buldyrev,
  Sciortino, and Tartaglia]{Zaccarelli2002}
E.~Zaccarelli, G.~Foffi, K.~A. Dawson, S.~V. Buldyrev, F.~Sciortino and
  P.~Tartaglia, \emph{Phys. Rev. E}, 2002, \textbf{66}, 041402\relax
\mciteBstWouldAddEndPuncttrue
\mciteSetBstMidEndSepPunct{\mcitedefaultmidpunct}
{\mcitedefaultendpunct}{\mcitedefaultseppunct}\relax
\EndOfBibitem
\bibitem[Pham \emph{et~al.}(2002)Pham, Puertas, Bergenholtz, Egelhaaf,
  Moussaıid, Pusey, Schofield, Cates, Fuchs, and Poon]{Pham2002}
K.~N. Pham, A.~M. Puertas, J.~Bergenholtz, S.~U. Egelhaaf, A.~Moussaıid, P.~N.
  Pusey, A.~B. Schofield, M.~E. Cates, M.~Fuchs and W.~C.~K. Poon,
  \emph{Science}, 2002, \textbf{296}, 104–106\relax
\mciteBstWouldAddEndPuncttrue
\mciteSetBstMidEndSepPunct{\mcitedefaultmidpunct}
{\mcitedefaultendpunct}{\mcitedefaultseppunct}\relax
\EndOfBibitem
\bibitem[Götze and Sperl(2003)]{WGotze_2003}
W.~Götze and M.~Sperl, \emph{Journal of Physics: Condensed Matter}, 2003,
  \textbf{15}, S869\relax
\mciteBstWouldAddEndPuncttrue
\mciteSetBstMidEndSepPunct{\mcitedefaultmidpunct}
{\mcitedefaultendpunct}{\mcitedefaultseppunct}\relax
\EndOfBibitem
\bibitem[Kaufman and Weitz(2006)]{Kaufman2006}
L.~J. Kaufman and D.~A. Weitz, \emph{The Journal of Chemical Physics}, 2006,
  \textbf{125}, 074716\relax
\mciteBstWouldAddEndPuncttrue
\mciteSetBstMidEndSepPunct{\mcitedefaultmidpunct}
{\mcitedefaultendpunct}{\mcitedefaultseppunct}\relax
\EndOfBibitem
\bibitem[Zaccarelli and Poon(2009)]{Zaccarelli2009}
E.~Zaccarelli and W.~C.~K. Poon, \emph{Proceedings of the National Academy of
  Sciences}, 2009, \textbf{106}, 15203–15208\relax
\mciteBstWouldAddEndPuncttrue
\mciteSetBstMidEndSepPunct{\mcitedefaultmidpunct}
{\mcitedefaultendpunct}{\mcitedefaultseppunct}\relax
\EndOfBibitem
\bibitem[Willenbacher \emph{et~al.}(2011)Willenbacher, Vesaratchanon,
  Thorwarth, and Bartsch]{Willenbacher2011}
N.~Willenbacher, J.~S. Vesaratchanon, O.~Thorwarth and E.~Bartsch, \emph{Soft
  Matter}, 2011, \textbf{7}, 5777--5788\relax
\mciteBstWouldAddEndPuncttrue
\mciteSetBstMidEndSepPunct{\mcitedefaultmidpunct}
{\mcitedefaultendpunct}{\mcitedefaultseppunct}\relax
\EndOfBibitem
\bibitem[Royall \emph{et~al.}(2018)Royall, Williams, and Tanaka]{Royall2018}
C.~P. Royall, S.~R. Williams and H.~Tanaka, \emph{The Journal of Chemical
  Physics}, 2018, \textbf{148}, 044501\relax
\mciteBstWouldAddEndPuncttrue
\mciteSetBstMidEndSepPunct{\mcitedefaultmidpunct}
{\mcitedefaultendpunct}{\mcitedefaultseppunct}\relax
\EndOfBibitem
\bibitem[Fullerton and Berthier(2020)]{Fullerton2020}
C.~J. Fullerton and L.~Berthier, \emph{Phys. Rev. Lett.}, 2020, \textbf{125},
  258004\relax
\mciteBstWouldAddEndPuncttrue
\mciteSetBstMidEndSepPunct{\mcitedefaultmidpunct}
{\mcitedefaultendpunct}{\mcitedefaultseppunct}\relax
\EndOfBibitem
\bibitem[Luo and Janssen(2021)]{Luo2021}
C.~Luo and L.~M.~C. Janssen, \emph{Soft Matter}, 2021, \textbf{17},
  7645--7661\relax
\mciteBstWouldAddEndPuncttrue
\mciteSetBstMidEndSepPunct{\mcitedefaultmidpunct}
{\mcitedefaultendpunct}{\mcitedefaultseppunct}\relax
\EndOfBibitem
\bibitem[Pham \emph{et~al.}(2008)Pham, Petekidis, Vlassopoulos, Egelhaaf, Poon,
  and Pusey]{Pham2008}
K.~N. Pham, G.~Petekidis, D.~Vlassopoulos, S.~U. Egelhaaf, W.~C.~K. Poon and
  P.~N. Pusey, \emph{Journal of Rheology}, 2008, \textbf{52}, 649–676\relax
\mciteBstWouldAddEndPuncttrue
\mciteSetBstMidEndSepPunct{\mcitedefaultmidpunct}
{\mcitedefaultendpunct}{\mcitedefaultseppunct}\relax
\EndOfBibitem
\bibitem[Atmuri \emph{et~al.}(2012)Atmuri, Peklaris, Kishore, and
  Bhatia]{Atmuri2012}
A.~K. Atmuri, G.~A. Peklaris, S.~Kishore and S.~R. Bhatia, \emph{Soft Matter},
  2012, \textbf{8}, 8965\relax
\mciteBstWouldAddEndPuncttrue
\mciteSetBstMidEndSepPunct{\mcitedefaultmidpunct}
{\mcitedefaultendpunct}{\mcitedefaultseppunct}\relax
\EndOfBibitem
\bibitem[Koumakis and Petekidis(2011)]{Koumakis2011}
N.~Koumakis and G.~Petekidis, \emph{Soft Matter}, 2011, \textbf{7}, 2456\relax
\mciteBstWouldAddEndPuncttrue
\mciteSetBstMidEndSepPunct{\mcitedefaultmidpunct}
{\mcitedefaultendpunct}{\mcitedefaultseppunct}\relax
\EndOfBibitem
\bibitem[Moghimi and Petekidis(2020)]{Moghimi2020}
E.~Moghimi and G.~Petekidis, \emph{Journal of Rheology}, 2020, \textbf{64},
  1209–1225\relax
\mciteBstWouldAddEndPuncttrue
\mciteSetBstMidEndSepPunct{\mcitedefaultmidpunct}
{\mcitedefaultendpunct}{\mcitedefaultseppunct}\relax
\EndOfBibitem
\bibitem[Laurati \emph{et~al.}(2009)Laurati, Petekidis, Koumakis, Cardinaux,
  Schofield, Brader, Fuchs, and Egelhaaf]{Laurati2009}
M.~Laurati, G.~Petekidis, N.~Koumakis, F.~Cardinaux, A.~B. Schofield, J.~M.
  Brader, M.~Fuchs and S.~U. Egelhaaf, \emph{The Journal of Chemical Physics},
  2009, \textbf{130}, 134907\relax
\mciteBstWouldAddEndPuncttrue
\mciteSetBstMidEndSepPunct{\mcitedefaultmidpunct}
{\mcitedefaultendpunct}{\mcitedefaultseppunct}\relax
\EndOfBibitem
\bibitem[Poon \emph{et~al.}(1993)Poon, Selfe, Robertson, Ilett, Pirie, and
  Pusey]{Poon1993}
W.~C.~K. Poon, J.~S. Selfe, M.~B. Robertson, S.~M. Ilett, A.~D. Pirie and P.~N.
  Pusey, \emph{Journal de Physique II}, 1993, \textbf{3}, 1075–1086\relax
\mciteBstWouldAddEndPuncttrue
\mciteSetBstMidEndSepPunct{\mcitedefaultmidpunct}
{\mcitedefaultendpunct}{\mcitedefaultseppunct}\relax
\EndOfBibitem
\bibitem[Eckert and Bartsch(2002)]{Eckert2002}
T.~Eckert and E.~Bartsch, \emph{Phys. Rev. Lett.}, 2002, \textbf{89},
  125701\relax
\mciteBstWouldAddEndPuncttrue
\mciteSetBstMidEndSepPunct{\mcitedefaultmidpunct}
{\mcitedefaultendpunct}{\mcitedefaultseppunct}\relax
\EndOfBibitem
\bibitem[Puertas \emph{et~al.}(2002)Puertas, Fuchs, and Cates]{Puertas2002}
A.~M. Puertas, M.~Fuchs and M.~E. Cates, \emph{Phys. Rev. Lett.}, 2002,
  \textbf{88}, 098301\relax
\mciteBstWouldAddEndPuncttrue
\mciteSetBstMidEndSepPunct{\mcitedefaultmidpunct}
{\mcitedefaultendpunct}{\mcitedefaultseppunct}\relax
\EndOfBibitem
\bibitem[Poon(2004)]{Poon2004}
W.~C. Poon, \emph{MRS Bulletin}, 2004, \textbf{29}, 96–99\relax
\mciteBstWouldAddEndPuncttrue
\mciteSetBstMidEndSepPunct{\mcitedefaultmidpunct}
{\mcitedefaultendpunct}{\mcitedefaultseppunct}\relax
\EndOfBibitem
\bibitem[Reichman and Charbonneau(2005)]{Reichman2005}
D.~R. Reichman and P.~Charbonneau, \emph{Journal of Statistical Mechanics:
  Theory and Experiment}, 2005, \textbf{2005}, P05013\relax
\mciteBstWouldAddEndPuncttrue
\mciteSetBstMidEndSepPunct{\mcitedefaultmidpunct}
{\mcitedefaultendpunct}{\mcitedefaultseppunct}\relax
\EndOfBibitem
\bibitem[Zaccarelli \emph{et~al.}(2008)Zaccarelli, Lu, Ciulla, Weitz, and
  Sciortino]{Zaccarelli2008}
E.~Zaccarelli, P.~J. Lu, F.~Ciulla, D.~A. Weitz and F.~Sciortino, \emph{Journal
  of Physics: Condensed Matter}, 2008, \textbf{20}, 494242\relax
\mciteBstWouldAddEndPuncttrue
\mciteSetBstMidEndSepPunct{\mcitedefaultmidpunct}
{\mcitedefaultendpunct}{\mcitedefaultseppunct}\relax
\EndOfBibitem
\bibitem[Puertas \emph{et~al.}(2002)Puertas, Fuchs, and Cates]{Cates2002}
A.~M. Puertas, M.~Fuchs and M.~E. Cates, \emph{Phys. Rev. Lett.}, 2002,
  \textbf{88}, 098301\relax
\mciteBstWouldAddEndPuncttrue
\mciteSetBstMidEndSepPunct{\mcitedefaultmidpunct}
{\mcitedefaultendpunct}{\mcitedefaultseppunct}\relax
\EndOfBibitem
\bibitem[Sciortino and Tartaglia(2005)]{Sciortino2005}
F.~Sciortino and P.~Tartaglia, \emph{Advances in Physics}, 2005, \textbf{54},
  471--524\relax
\mciteBstWouldAddEndPuncttrue
\mciteSetBstMidEndSepPunct{\mcitedefaultmidpunct}
{\mcitedefaultendpunct}{\mcitedefaultseppunct}\relax
\EndOfBibitem
\bibitem[Berthier and Tarjus(2009)]{Berthier2009}
L.~Berthier and G.~Tarjus, \emph{Phys. Rev. Lett.}, 2009, \textbf{103},
  170601\relax
\mciteBstWouldAddEndPuncttrue
\mciteSetBstMidEndSepPunct{\mcitedefaultmidpunct}
{\mcitedefaultendpunct}{\mcitedefaultseppunct}\relax
\EndOfBibitem
\bibitem[Schweizer(2005)]{Schweizer2005}
K.~S. Schweizer, \emph{The Journal of Chemical Physics}, 2005, \textbf{123},
  244501\relax
\mciteBstWouldAddEndPuncttrue
\mciteSetBstMidEndSepPunct{\mcitedefaultmidpunct}
{\mcitedefaultendpunct}{\mcitedefaultseppunct}\relax
\EndOfBibitem
\bibitem[Mayer \emph{et~al.}(2006)Mayer, Miyazaki, and Reichman]{Mayer2006}
P.~Mayer, K.~Miyazaki and D.~R. Reichman, \emph{Phys. Rev. Lett.}, 2006,
  \textbf{97}, 095702\relax
\mciteBstWouldAddEndPuncttrue
\mciteSetBstMidEndSepPunct{\mcitedefaultmidpunct}
{\mcitedefaultendpunct}{\mcitedefaultseppunct}\relax
\EndOfBibitem
\bibitem[Janssen and Reichman(2015)]{Janssen2015}
L.~M.~C. Janssen and D.~R. Reichman, \emph{Phys. Rev. Lett.}, 2015,
  \textbf{115}, 205701\relax
\mciteBstWouldAddEndPuncttrue
\mciteSetBstMidEndSepPunct{\mcitedefaultmidpunct}
{\mcitedefaultendpunct}{\mcitedefaultseppunct}\relax
\EndOfBibitem
\bibitem[Mirigian and Schweizer(2014)]{Mirigian2014}
S.~Mirigian and K.~S. Schweizer, \emph{The Journal of Chemical Physics}, 2014,
  \textbf{140}, 194506\relax
\mciteBstWouldAddEndPuncttrue
\mciteSetBstMidEndSepPunct{\mcitedefaultmidpunct}
{\mcitedefaultendpunct}{\mcitedefaultseppunct}\relax
\EndOfBibitem
\bibitem[Mirigian and Schweizer(2013)]{Mirigian2013}
S.~Mirigian and K.~S. Schweizer, \emph{The Journal of Physical Chemistry
  Letters}, 2013, \textbf{4}, 3648–3653\relax
\mciteBstWouldAddEndPuncttrue
\mciteSetBstMidEndSepPunct{\mcitedefaultmidpunct}
{\mcitedefaultendpunct}{\mcitedefaultseppunct}\relax
\EndOfBibitem
\bibitem[Dell and Schweizer(2015)]{Dell2015}
Z.~E. Dell and K.~S. Schweizer, \emph{Phys. Rev. Lett.}, 2015, \textbf{115},
  205702\relax
\mciteBstWouldAddEndPuncttrue
\mciteSetBstMidEndSepPunct{\mcitedefaultmidpunct}
{\mcitedefaultendpunct}{\mcitedefaultseppunct}\relax
\EndOfBibitem
\bibitem[Ghosh and Schweizer(2019)]{Ghosh2019}
A.~Ghosh and K.~S. Schweizer, \emph{The Journal of Chemical Physics}, 2019,
  \textbf{151}, 244502\relax
\mciteBstWouldAddEndPuncttrue
\mciteSetBstMidEndSepPunct{\mcitedefaultmidpunct}
{\mcitedefaultendpunct}{\mcitedefaultseppunct}\relax
\EndOfBibitem
\bibitem[Ghosh and Schweizer(2020)]{Ghosh2020}
A.~Ghosh and K.~S. Schweizer, \emph{Phys. Rev. E}, 2020, \textbf{101},
  060601\relax
\mciteBstWouldAddEndPuncttrue
\mciteSetBstMidEndSepPunct{\mcitedefaultmidpunct}
{\mcitedefaultendpunct}{\mcitedefaultseppunct}\relax
\EndOfBibitem
\bibitem[Berthier and Tarjus(2010)]{Berthier2010}
L.~Berthier and G.~Tarjus, \emph{Phys. Rev. E}, 2010, \textbf{82}, 031502\relax
\mciteBstWouldAddEndPuncttrue
\mciteSetBstMidEndSepPunct{\mcitedefaultmidpunct}
{\mcitedefaultendpunct}{\mcitedefaultseppunct}\relax
\EndOfBibitem
\bibitem[Tong and Tanaka(2020)]{Tong2020}
H.~Tong and H.~Tanaka, \emph{Phys. Rev. Lett.}, 2020, \textbf{124},
  225501\relax
\mciteBstWouldAddEndPuncttrue
\mciteSetBstMidEndSepPunct{\mcitedefaultmidpunct}
{\mcitedefaultendpunct}{\mcitedefaultseppunct}\relax
\EndOfBibitem
\bibitem[Landes \emph{et~al.}(2020)Landes, Biroli, Dauchot, Liu, and
  Reichman]{Landes2020}
F.~m. c.~P. Landes, G.~Biroli, O.~Dauchot, A.~J. Liu and D.~R. Reichman,
  \emph{Phys. Rev. E}, 2020, \textbf{101}, 010602\relax
\mciteBstWouldAddEndPuncttrue
\mciteSetBstMidEndSepPunct{\mcitedefaultmidpunct}
{\mcitedefaultendpunct}{\mcitedefaultseppunct}\relax
\EndOfBibitem
\bibitem[Nandi \emph{et~al.}(2017)Nandi, Banerjee, Dasgupta, and
  Bhattacharyya]{Nandi2017}
M.~K. Nandi, A.~Banerjee, C.~Dasgupta and S.~M. Bhattacharyya, \emph{Phys. Rev.
  Lett.}, 2017, \textbf{119}, 265502\relax
\mciteBstWouldAddEndPuncttrue
\mciteSetBstMidEndSepPunct{\mcitedefaultmidpunct}
{\mcitedefaultendpunct}{\mcitedefaultseppunct}\relax
\EndOfBibitem
\bibitem[Berthier and Tarjus(2011)]{Berthier2011_2}
L.~Berthier and G.~Tarjus, \emph{The Journal of Chemical Physics}, 2011,
  \textbf{134}, 214503\relax
\mciteBstWouldAddEndPuncttrue
\mciteSetBstMidEndSepPunct{\mcitedefaultmidpunct}
{\mcitedefaultendpunct}{\mcitedefaultseppunct}\relax
\EndOfBibitem
\bibitem[Coslovich and Pastore(2007)]{Coslovich2007}
D.~Coslovich and G.~Pastore, \emph{The Journal of Chemical Physics}, 2007,
  \textbf{127}, 124504\relax
\mciteBstWouldAddEndPuncttrue
\mciteSetBstMidEndSepPunct{\mcitedefaultmidpunct}
{\mcitedefaultendpunct}{\mcitedefaultseppunct}\relax
\EndOfBibitem
\bibitem[Chen \emph{et~al.}(2010)Chen, Ellenbroek, Zhang, Chen, Yunker, Henkes,
  Brito, Dauchot, van Saarloos, Liu, and Yodh]{Chen2010}
K.~Chen, W.~G. Ellenbroek, Z.~Zhang, D.~T.~N. Chen, P.~J. Yunker, S.~Henkes,
  C.~Brito, O.~Dauchot, W.~van Saarloos, A.~J. Liu and A.~G. Yodh, \emph{Phys.
  Rev. Lett.}, 2010, \textbf{105}, 025501\relax
\mciteBstWouldAddEndPuncttrue
\mciteSetBstMidEndSepPunct{\mcitedefaultmidpunct}
{\mcitedefaultendpunct}{\mcitedefaultseppunct}\relax
\EndOfBibitem
\bibitem[Zylberg \emph{et~al.}(2017)Zylberg, Lerner, Bar-Sinai, and
  Bouchbinder]{Jacques2017}
J.~Zylberg, E.~Lerner, Y.~Bar-Sinai and E.~Bouchbinder, \emph{Proceedings of
  the National Academy of Sciences}, 2017, \textbf{114}, 7289--7294\relax
\mciteBstWouldAddEndPuncttrue
\mciteSetBstMidEndSepPunct{\mcitedefaultmidpunct}
{\mcitedefaultendpunct}{\mcitedefaultseppunct}\relax
\EndOfBibitem
\bibitem[Schweizer and Chandler(1982)]{Schweizer1982}
K.~S. Schweizer and D.~Chandler, \emph{The Journal of Chemical Physics}, 1982,
  \textbf{76}, 2296--2314\relax
\mciteBstWouldAddEndPuncttrue
\mciteSetBstMidEndSepPunct{\mcitedefaultmidpunct}
{\mcitedefaultendpunct}{\mcitedefaultseppunct}\relax
\EndOfBibitem
\bibitem[Schweizer(1989)]{Schweizer1989}
K.~S. Schweizer, \emph{The Journal of Chemical Physics}, 1989, \textbf{91},
  5802--5821\relax
\mciteBstWouldAddEndPuncttrue
\mciteSetBstMidEndSepPunct{\mcitedefaultmidpunct}
{\mcitedefaultendpunct}{\mcitedefaultseppunct}\relax
\EndOfBibitem
\bibitem[Hansen and McDonald(2006)]{Hansen2006-pc}
J.~P. Hansen and I.~R. McDonald, \emph{Theory of simple liquids}, Academic
  Press Inc. (London), London, England, 3rd edn, 2006\relax
\mciteBstWouldAddEndPuncttrue
\mciteSetBstMidEndSepPunct{\mcitedefaultmidpunct}
{\mcitedefaultendpunct}{\mcitedefaultseppunct}\relax
\EndOfBibitem
\bibitem[Ghosh and Schweizer(2020)]{Ghosh2020_2}
A.~Ghosh and K.~S. Schweizer, \emph{The Journal of Chemical Physics}, 2020,
  \textbf{153}, 194502\relax
\mciteBstWouldAddEndPuncttrue
\mciteSetBstMidEndSepPunct{\mcitedefaultmidpunct}
{\mcitedefaultendpunct}{\mcitedefaultseppunct}\relax
\EndOfBibitem
\bibitem[Ghosh(2023)]{Ghosh2023}
A.~Ghosh, \emph{The Journal of Physical Chemistry B}, 2023, \textbf{127},
  5162–5168\relax
\mciteBstWouldAddEndPuncttrue
\mciteSetBstMidEndSepPunct{\mcitedefaultmidpunct}
{\mcitedefaultendpunct}{\mcitedefaultseppunct}\relax
\EndOfBibitem
\bibitem[Schweizer and Saltzman(2003)]{Saltzman2003}
K.~S. Schweizer and E.~J. Saltzman, \emph{The Journal of Chemical Physics},
  2003, \textbf{119}, 1181--1196\relax
\mciteBstWouldAddEndPuncttrue
\mciteSetBstMidEndSepPunct{\mcitedefaultmidpunct}
{\mcitedefaultendpunct}{\mcitedefaultseppunct}\relax
\EndOfBibitem
\bibitem[Kirkpatrick and Wolynes(1987)]{Kirkpatrick1987}
T.~R. Kirkpatrick and P.~G. Wolynes, \emph{Phys. Rev. A}, 1987, \textbf{35},
  3072--3080\relax
\mciteBstWouldAddEndPuncttrue
\mciteSetBstMidEndSepPunct{\mcitedefaultmidpunct}
{\mcitedefaultendpunct}{\mcitedefaultseppunct}\relax
\EndOfBibitem
\bibitem[Verlet(1980)]{Verlet1980}
L.~Verlet, \emph{Molecular Physics}, 1980, \textbf{41}, 183–190\relax
\mciteBstWouldAddEndPuncttrue
\mciteSetBstMidEndSepPunct{\mcitedefaultmidpunct}
{\mcitedefaultendpunct}{\mcitedefaultseppunct}\relax
\EndOfBibitem
\bibitem[Zhou \emph{et~al.}(2020)Zhou, Mei, and Schweizer]{Zhou2020}
Y.~Zhou, B.~Mei and K.~S. Schweizer, \emph{Phys. Rev. E}, 2020, \textbf{101},
  042121\relax
\mciteBstWouldAddEndPuncttrue
\mciteSetBstMidEndSepPunct{\mcitedefaultmidpunct}
{\mcitedefaultendpunct}{\mcitedefaultseppunct}\relax
\EndOfBibitem
\bibitem[N\"{a}gele and Bergenholtz(1998)]{Ngele1998}
G.~N\"{a}gele and J.~Bergenholtz, \emph{The Journal of Chemical Physics}, 1998,
  \textbf{108}, 9893–9904\relax
\mciteBstWouldAddEndPuncttrue
\mciteSetBstMidEndSepPunct{\mcitedefaultmidpunct}
{\mcitedefaultendpunct}{\mcitedefaultseppunct}\relax
\EndOfBibitem
\bibitem[Dyre \emph{et~al.}(1996)Dyre, Olsen, and Christensen]{Dyre1996}
J.~C. Dyre, N.~B. Olsen and T.~Christensen, \emph{Phys. Rev. B}, 1996,
  \textbf{53}, 2171--2174\relax
\mciteBstWouldAddEndPuncttrue
\mciteSetBstMidEndSepPunct{\mcitedefaultmidpunct}
{\mcitedefaultendpunct}{\mcitedefaultseppunct}\relax
\EndOfBibitem
\bibitem[Dyre(1998)]{Dyre1998}
J.~C. Dyre, \emph{Journal of Non-Crystalline Solids}, 1998, \textbf{235-237},
  142--149\relax
\mciteBstWouldAddEndPuncttrue
\mciteSetBstMidEndSepPunct{\mcitedefaultmidpunct}
{\mcitedefaultendpunct}{\mcitedefaultseppunct}\relax
\EndOfBibitem
\bibitem[Dyre \emph{et~al.}(2006)Dyre, Christensen, and Olsen]{Dyre2006}
J.~C. Dyre, T.~Christensen and N.~B. Olsen, \emph{Journal of Non-Crystalline
  Solids}, 2006, \textbf{352}, 4635--4642\relax
\mciteBstWouldAddEndPuncttrue
\mciteSetBstMidEndSepPunct{\mcitedefaultmidpunct}
{\mcitedefaultendpunct}{\mcitedefaultseppunct}\relax
\EndOfBibitem
\bibitem[Xie and Schweizer(2016)]{Xie2016}
S.-J. Xie and K.~S. Schweizer, \emph{Macromolecules}, 2016, \textbf{49},
  9655–9664\relax
\mciteBstWouldAddEndPuncttrue
\mciteSetBstMidEndSepPunct{\mcitedefaultmidpunct}
{\mcitedefaultendpunct}{\mcitedefaultseppunct}\relax
\EndOfBibitem
\bibitem[Mei \emph{et~al.}(2020)Mei, Zhou, and Schweizer]{Mei2020}
B.~Mei, Y.~Zhou and K.~S. Schweizer, \emph{The Journal of Physical Chemistry
  B}, 2020, \textbf{124}, 6121–6131\relax
\mciteBstWouldAddEndPuncttrue
\mciteSetBstMidEndSepPunct{\mcitedefaultmidpunct}
{\mcitedefaultendpunct}{\mcitedefaultseppunct}\relax
\EndOfBibitem
\bibitem[Zhou and Schweizer(2019)]{Zhou2019}
Y.~Zhou and K.~S. Schweizer, \emph{The Journal of Chemical Physics}, 2019,
  \textbf{150}, 214902\relax
\mciteBstWouldAddEndPuncttrue
\mciteSetBstMidEndSepPunct{\mcitedefaultmidpunct}
{\mcitedefaultendpunct}{\mcitedefaultseppunct}\relax
\EndOfBibitem
\bibitem[Priya and Voigtmann(2014)]{Priya2014}
M.~Priya and T.~Voigtmann, \emph{Journal of Rheology}, 2014, \textbf{58},
  1163--1187\relax
\mciteBstWouldAddEndPuncttrue
\mciteSetBstMidEndSepPunct{\mcitedefaultmidpunct}
{\mcitedefaultendpunct}{\mcitedefaultseppunct}\relax
\EndOfBibitem
\bibitem[Chandler \emph{et~al.}(1983)Chandler, Weeks, and
  Andersen]{Chandler1983}
D.~Chandler, J.~D. Weeks and H.~C. Andersen, \emph{Science}, 1983,
  \textbf{220}, 787--794\relax
\mciteBstWouldAddEndPuncttrue
\mciteSetBstMidEndSepPunct{\mcitedefaultmidpunct}
{\mcitedefaultendpunct}{\mcitedefaultseppunct}\relax
\EndOfBibitem
\bibitem[Kobelev and Schweizer(2005)]{Kobelev2005}
V.~Kobelev and K.~S. Schweizer, \emph{Phys. Rev. E}, 2005, \textbf{71},
  021401\relax
\mciteBstWouldAddEndPuncttrue
\mciteSetBstMidEndSepPunct{\mcitedefaultmidpunct}
{\mcitedefaultendpunct}{\mcitedefaultseppunct}\relax
\EndOfBibitem
\bibitem[Rao \emph{et~al.}(2006)Rao, Kobelev, Li, Lewis, and
  Schweizer]{Rao2006}
R.~B. Rao, V.~L. Kobelev, Q.~Li, J.~A. Lewis and K.~S. Schweizer,
  \emph{Langmuir}, 2006, \textbf{22}, 2441–2443\relax
\mciteBstWouldAddEndPuncttrue
\mciteSetBstMidEndSepPunct{\mcitedefaultmidpunct}
{\mcitedefaultendpunct}{\mcitedefaultseppunct}\relax
\EndOfBibitem
\bibitem[Chen and Schweizer(2004)]{Chen2004}
Y.-L. Chen and K.~S. Schweizer, \emph{The Journal of Chemical Physics}, 2004,
  \textbf{120}, 7212–7222\relax
\mciteBstWouldAddEndPuncttrue
\mciteSetBstMidEndSepPunct{\mcitedefaultmidpunct}
{\mcitedefaultendpunct}{\mcitedefaultseppunct}\relax
\EndOfBibitem
\bibitem[Chen \emph{et~al.}(2005)Chen, Kobelev, and Schweizer]{Chen2005}
Y.-L. Chen, V.~Kobelev and K.~S. Schweizer, \emph{Phys. Rev. E}, 2005,
  \textbf{71}, 041405\relax
\mciteBstWouldAddEndPuncttrue
\mciteSetBstMidEndSepPunct{\mcitedefaultmidpunct}
{\mcitedefaultendpunct}{\mcitedefaultseppunct}\relax
\EndOfBibitem
\bibitem[Asakura and Oosawa(1954)]{Asakura1954}
S.~Asakura and F.~Oosawa, \emph{The Journal of Chemical Physics}, 1954,
  \textbf{22}, 1255--1256\relax
\mciteBstWouldAddEndPuncttrue
\mciteSetBstMidEndSepPunct{\mcitedefaultmidpunct}
{\mcitedefaultendpunct}{\mcitedefaultseppunct}\relax
\EndOfBibitem
\bibitem[Miyazaki \emph{et~al.}(2022)Miyazaki, Schweizer, Thirumalai, Tuinier,
  and Zaccarelli]{AOPotentialJCP2022}
K.~Miyazaki, K.~S. Schweizer, D.~Thirumalai, R.~Tuinier and E.~Zaccarelli,
  \emph{The Journal of Chemical Physics}, 2022, \textbf{156}, 080401\relax
\mciteBstWouldAddEndPuncttrue
\mciteSetBstMidEndSepPunct{\mcitedefaultmidpunct}
{\mcitedefaultendpunct}{\mcitedefaultseppunct}\relax
\EndOfBibitem
\bibitem[Schweizer and Saltzman(2004)]{Schweizer2004}
K.~S. Schweizer and E.~J. Saltzman, \emph{The Journal of Physical Chemistry B},
  2004, \textbf{108}, 19729–19741\relax
\mciteBstWouldAddEndPuncttrue
\mciteSetBstMidEndSepPunct{\mcitedefaultmidpunct}
{\mcitedefaultendpunct}{\mcitedefaultseppunct}\relax
\EndOfBibitem
\bibitem[Xie and Schweizer(2020)]{Xie2020}
S.-J. Xie and K.~S. Schweizer, \emph{The Journal of Chemical Physics}, 2020,
  \textbf{152}, 034502\relax
\mciteBstWouldAddEndPuncttrue
\mciteSetBstMidEndSepPunct{\mcitedefaultmidpunct}
{\mcitedefaultendpunct}{\mcitedefaultseppunct}\relax
\EndOfBibitem
\bibitem[Ghosh and Schweizer(2023)]{Ghosh2023_2}
A.~Ghosh and K.~S. Schweizer, \emph{Journal of Rheology}, 2023, \textbf{67},
  559--578\relax
\mciteBstWouldAddEndPuncttrue
\mciteSetBstMidEndSepPunct{\mcitedefaultmidpunct}
{\mcitedefaultendpunct}{\mcitedefaultseppunct}\relax
\EndOfBibitem
\bibitem[Koumakis and Petekidis(2011)]{Kaumakis2011}
N.~Koumakis and G.~Petekidis, \emph{Soft Matter}, 2011, \textbf{7},
  2456--2470\relax
\mciteBstWouldAddEndPuncttrue
\mciteSetBstMidEndSepPunct{\mcitedefaultmidpunct}
{\mcitedefaultendpunct}{\mcitedefaultseppunct}\relax
\EndOfBibitem
\end{mcitethebibliography}
\bibliographystyle{rsc} 

\end{document}